\documentclass[onecolumn]{IEEEtran}
\usepackage{times}

\usepackage[top=0.6in, bottom=0.7in, left=0.6in, right=0.6in]{geometry}

\usepackage{amsmath}  
\usepackage{amssymb}  
\usepackage{mathrsfs} 

\usepackage{theorem}  
\usepackage{cite}     
\usepackage{comment}  

\usepackage{upref}
\usepackage{amsfonts}

\usepackage{verbatim}

\usepackage[dvipsnames,usenames]{color}
\usepackage{enumerate}

\usepackage{graphicx}
\usepackage{subfigure}

\usepackage{latexsym}

\usepackage[draft,ulem=normalem]{changes}
\usepackage[all]{xy}

\newcommand{\be}[1]{\begin{equation}\label{#1}}
\newcommand{\ee}{\end{equation}}

\newcommand{\bc}{\begin{center}}
\newcommand{\ec}{\end{center}}

\newcommand{\cB}{{\cal B}}
\newcommand{\cC}{{\cal C}}
\newcommand{\cD}{{\cal D}}

\newcommand{\cO}{{\cal O}}

\newcommand{\cX}{{\cal X}}
\newcommand{\cY}{{\cal Y}}

\newcommand{\bfa}{{\boldsymbol a}}

\newcommand{\bfc}{{\boldsymbol c}}
\newcommand{\bfd}{{\boldsymbol d}}
\newcommand{\bfe}{{\boldsymbol e}}

\newcommand{\bfs}{{\boldsymbol s}}

\newcommand{\bfu}{{\boldsymbol u}}
\newcommand{\bfv}{{\boldsymbol v}}
\newcommand{\bfw}{{\boldsymbol w}}
\newcommand{\bfx}{{\boldsymbol x}}
\newcommand{\bfy}{{\boldsymbol y}}
\newcommand{\bfz}{{\boldsymbol z}}

\newcommand{\bfC}{{\mathbf C}}
\newcommand{\bfD}{{\mathbf D}}

\renewcommand{\leq}{\leqslant}

\renewcommand{\geq}{\geqslant}

\newcommand{\Cref}[1]{Co\-rol\-la\-ry\,\ref{#1}}

\theoremstyle{plain} \theorembodyfont{\normalfont\slshape}

\newtheorem{thm}{Theorem$\!$}
\newenvironment{theorem}{\begin{thm}\hspace*{-1ex}{\bf.}}{\end{thm}}

\newtheorem{prop}[thm]{Proposition$\!$}

\newtheorem{lem}[thm]{Lemma$\!$}
\newenvironment{lemma}{\begin{lem}\hspace*{-1ex}{\bf.}}{\end{lem}}

\newtheorem{cor}[thm]{Corollary$\!$}
\newenvironment{corollary}{\begin{cor}\hspace*{-1ex}{\bf.}}{\end{cor}}

\newtheorem{prob}[thm]{Problem$\!$}

\newtheorem{defi}[thm]{Definition$\!$}
\newenvironment{definition}{\begin{defi}\hspace*{-1ex}{\bf.}}{\end{defi}}

\newtheorem{claims}{Claim$\!$}
\newenvironment{claim}{\begin{claims}\hspace*{-1ex}{\bf .}}{\end{claims}}

\theorembodyfont{\normalfont}

\newtheorem{exam}{Example$\!$}
\newenvironment{example}{\begin{exam}\hspace*{-1ex}{\bf .}}{\end{exam}}

\newtheorem{remrk}{Remark$\!$}
\newenvironment{remark}{\begin{remrk}\hspace*{-1ex}{\bf .}}{\end{remrk}}

\definecolor{Codecolor}{named}{White}

\newcommand{\Copen}{\mbox{\{\kern-5.50pt\{}}
\newcommand{\Cclose}{\mbox{\}\kern-5.50pt\}}}
\newcommand{\Cslash}{\mbox{$\backslash\kern-6.02pt\backslash$}}

\begin{document}

\title{Codes in the Damerau Distance for Deletion and Adjacent Transposition Correction}
\author{
  \IEEEauthorblockN{
    Ryan~Gabrys\IEEEauthorrefmark{1},~ 
    Eitan~Yaakobi~\IEEEauthorrefmark{3},~and
    Olgica~Milenkovic\IEEEauthorrefmark{1}}
  {\normalsize
    \begin{tabular}{cc}
      \IEEEauthorrefmark{1}ECE Department, University of Illinois, Urbana-Champaign & \IEEEauthorrefmark{3}Technion University\\
           
    \end{tabular}}
   
    }
    
\maketitle
\begin{abstract} Motivated by applications in DNA-based storage, we introduce the new problem of code design in the Damerau metric. The Damerau metric is a generalization of the Levenshtein distance which, in addition to deletions, insertions and substitution errors also accounts for adjacent transposition edits. We first provide constructions for codes that may correct either a single deletion or a single adjacent transposition and then proceed to extend these results to codes that can simultaneously correct a single deletion and multiple adjacent transpositions. We conclude with constructions for joint block deletion and adjacent block transposition error-correcting codes.\footnote{Parts of the results were presented at the International Symposium on Information Theory in Barcelona, 2016.}
\end{abstract}

\section{Introduction} \label{sec:intro}

The edit distance is a measure of similarity between two strings evaluated based on the minimum number of operations required to transform one string into the other. If the operations are confined to symbol deletions, insertions and substitutions, the distance of interest is the Levenshtein (edit) distance~\cite{levenshtein1966binary}. The Levenshtein distance has found numerous applications in bioinformatics, where a weighted version of this metric is used to assess the similarity of DNA strings and reconstruct phylogenetic trees~\cite{kumar2004mega3}, and natural language processing, where the distance is used to model spelling errors and provide automated word correction~\cite{brill2000improved}.

In parallel to the work on developing efficient algorithms for computing the edit distance and performing alignments of large number of strings, a long line of results were reported on the topic of designing codes for this distance function. Codes in the edit distance are of particular importance for communication in the presence of synchronization errors, a type of error encountered in almost all modern storage and data transmission systems. Classical derivations of upper bounds on code sizes by Levenshtein~\cite{levenshtein1966binary} and single deletion-correcting code constructions by Varshamov and Tenengoltz~\cite{varshamov1965code,sloane2002single} have established the framework for studying many challenging problems in optimal code design for this metric~\cite{schulman1999asymptotically,helberg2002multiple,cullina2014improvement,sala2016exact,brakensiek2015efficient}.
 
The Damerau distance is an extension of the Levenshtein distance that also allows for edits of the form of adjacent symbol transpositions~\cite{brill2000improved}. Despite the apparent interest in coding for edit channels, the problem of designing codes in the Damerau distance was not studied before. A possible reason for this lack of interest in the Damerau distance may be attributed to the fact that not many practical channel models involve adjacent transposition errors, and even if they do so, they tend not to allow for user-selected message\footnote{We note the an adjacent transposition may be viewed as a deletion/insertion pair. However, the locations of the deletion and insertion are adjacent, and hence correlated -- correcting for two random indel errors is in this case suboptimal. Codes in the Damerau distance address this problem by handling a combination of random deletions and correlated (adjacent) indels.}. Our motivating application for studying codes in the Damerau distance is the emerging paradigm of DNA-based storage~\cite{church2012next,goldman2013towards,yazdi2014dna,yazdi2015dna,bornholt2016dna,yazdi2017dna}. In DNA-based storage systems, media degradation arises due to DNA aging caused by metabolic and hydrolitic processes, or more precisely, by exposure to standard or increased level radiation, humidity, and high temperatures. As an example, human cellular DNA undergoes anywhere between 10-50 breakages in a cell cycle~\cite{vilenchik2003endogenous}. These DNA breakages or symbol/block deletions result in changed structures of the string: If a string breaks in two places, which is the most likely scenario, either the sequence reattaches itself without resulting in structural damage, reattaches itself in the opposite direction, resulting in what is called a \emph{reversal error}, or the broken string degrades, resulting in a bursty (block) deletion; if a string breaks in three positions, which is the second most likely breakage scenario, either the adjacent broken blocks exchange positions or one or both block disintegrate leading to a bursty deletion. It is the latter scenario that motivates the study of channels in which adjacent blocks of symbols may be exchanges or individual blocks deleted. It is straightforward to see that this editing scenario corresponds to a ``block version'' of the Damerau editing process. The block editing process is hard to analyze directly, so we first study the symbol-level Damerau editing process and then proceed to analyze the block model. Also, for simplicity of exposition, we focus our attention on deletion and adjacent transposition errors and delegate the more complex analysis of all four edit operations to future work.

Our contributions are two-fold. We introduce the Damerau distance code design problem, and describe the first known scheme for correcting one deletion \emph{or} one adjacent transposition. The scheme has near-optimal redundancy. We then proceed to extend and generalize this construction so as to obtain codes capable of correcting one deletion \emph{and} one adjacent transposition that also have near-optimal redundancy. Our results also shed light on the new problems of \emph{mismatched} Varshamov-Tenengoltz (VT) decoding and run length limited VT codes.  Second, we describe significantly more involved code constructions for correction of multiple adjacent transposition errors and proceed to introduce codes capable of correcting a block deletion and adjacent block transposition. In the derivation process, we improve upon the best known constructions for block deletion-correcting codes (i.e., codes capable of correcting a block of consecutive deletions).

The paper is organized as follows. Section~\ref{sec:notation} contains the problem statement and relevant notation. Section~\ref{sec:basics} contains an analysis of the code design procedure for single deletion or single adjacent transposition correction. Section~\ref{sec:main} contains an order optimal code construction for correcting a single deletion and a single adjacent transposition, as well a low-redundancy construction for codes correcting a single deletion and multiple adjacent transpositions. Sections~\ref{sec:block} and~\ref{sec:blockTransDel} are devoted to our main findings: The best known code construction for single block deletion correction, and codes capable of correcting a single block deletion and a single adjacent block transposition.

\section{Terminology and Notation}\label{sec:notation}

We start by defining the \emph{Damerau-Levenshtein distance}, which arose in the works of Damerau~\cite{damerau1964technique} and Levenshtein~\cite{levenshtein1966binary}, and by introducing codes in this metric. We then proceed to extend the underlying coding problem so that it applies to blocks, rather than individual symbol errors.

\begin{definition}
The \textbf{Damerau--Levenshtein distance} is a string metric, which for two strings of possibly different lengths over some (finite) alphabet equals the minimum number of insertions, deletions, substitutions and adjacent transposition edits needed to transform one string into the other. The \textbf{block Damerau--Levenshtein distance} with block length $b$ is a string metric, which for two strings of possibly different lengths over some (finite) alphabet equals the minimum number of insertions, deletions, substitutions and adjacent transposition edits of blocks of length at most $b$ needed to transform one string into the other.
\end{definition}

For simplicity, we focus on edits involving deletions and adjacent transpositions only, and with slight abuse of terminology refer to the underlying sequence comparison function as the Damerau metric\footnote{Since we only consider deletions, what we refer to as Damerau distance is strictly speaking not a metric, but we use the terminology as it is custom to do so.}. Furthermore, we restrict our attention to binary alphabets only. Generalizations to larger alphabet sizes may potentially be accomplished by a careful use of Tenegoltz up-down encoding, described in~\cite{paluncic2011note,le2015new}, but this problem will be discussed elsewhere. 

For a vector $\bfx \in \mathbb{F}_2^n$, let $\cB_{T \lor D}(\bfx)$ denote the set of vectors that may be obtained from $\bfx$ by either at most one single adjacent transposition (T) \emph{or} at most one single deletion (D). Note that the size of $\cB_{T \lor D}(\bfx)$ is $2r(\bfx)$, where $r(\bfx)$ is the number of runs in $\bfx$, i.e., the smallest number of nonoverlapping substrings involving the same symbol that ``covers'' the sequence. 
\begin{example} Suppose that $\bfx=(0,0,1,1,0) \in \mathbb{F}_2^n$. Then,
\begin{align}
\cB_{T \lor D}(\bfx) = \{ &(0,1,1,0), (0,0,1,0), (0,0,1,1), \notag \\
&(0,0,1,1,0),(0,1,0,1,0), (0,0,1,0,1) \}. \notag
\end{align}
In particular, $\cB_{T \lor D}(\bfx)=\cB_D(\bfx) \cup \cB_T(\bfx)$, where $\cB_D(\bfx)$ is the set of words obtained by deleting at most one element in $\bfx$, while
$\cB_T(\bfx)$ is the set of words obtained from at most one adjacent transposition in $\bfx$.
\end{example} 

The derivative of $\bfx$, denoted by $\partial(\bfx)=\bfx'$ is a vector defined as $\bfx' = (x_1,x_2+x_1,x_3+x_2,\ldots, x_n+x_{n-1})$. Clearly, the mapping between $\bfx$ and $\bfx'$ is a bijection. Hence, the integral $\partial^{-1}(\bfx)\triangleq  \overline{\bfx}$ is well-defined for all $\bfx\in \mathbb{F}_2^n$. Observe that $\partial^{-1}(\bfx) = (\bar{x}_1, \bar{x}_2, \ldots, \bar{x}_n) \in \mathbb{F}^n_2$, where $\bar{x}_i = \sum_{j=1}^i x_j$ for all $i \in [n]$. For a set $\cX \subseteq \mathbb{F}_2^n$, we use $\cX'$ to denote the set of derivatives of vectors in $\cX$, and similarly, we use $\overline{\cX}$ to denote the set of integrals of 
vectors in $\cX$. For two vectors $\bfx, \bfy \in \mathbb{F}_2^n$, we let $d_H(\bfx, \bfy)$ denote their Hamming distance. Furthermore, we let $\cC_H(n,d)$ stand for any code of length $n$ with minimum Hamming distance $d$, and similarly, we let $\cC_{D}(n)$ stand for any single-deletion-correcting code of length $n$.

\begin{table*}[t!]\renewcommand{\arraystretch}{1.5}\label{tab:notations}
\begin{center}
\begin{tabular}{ c c c }
 Notation & Description & Position in the manuscript \\ 
\hline
$\cB_{D}(\bfx)$ & The set of words that may be obtained from at most one single deletion in a vector $\bfx$. & End of Section~\ref{sec:notation}. \\
$\cB_{T}(\bfx)$ & The set of words that may be obtained from at most one single adjacent transposition in a vector $\bfx$. & End of Section~\ref{sec:notation}. \\
$\cB_{T \lor D}(\bfx)$ & $\cB_{T \lor D}(\bfx) = \cB_{D}(\bfx) \cup \cB_{T}(\bfx).$ & End of Section~\ref{sec:notation}. \\
$\bfx'$, $\partial(\bfx)$ & The derivative of $\bfx$. & End of Section~\ref{sec:notation}. \\  
$\bar{\bfx}$, $\partial^{-1}(\bfx)$  & The integral of $\bfx$ & End of Section~\ref{sec:notation}.\\
$\cC_H(n,d)$ & A code of minimum Hamming distance $d$. & End of Section~\ref{sec:notation}.\\
$\cC_D(n)$ & A code that can correct a single deletion error. & End of Section~\ref{sec:notation}.\\
\hline
$\cC_{T \lor D}(n)$ & A code that can correct a single adjacent transposition or deletion. & Section~\ref{sec:basics},  preceding Lemma~\ref{lem:t1}.\\
$\textbf{X}_D(n,a)$ & A code that can correct a single deletion error. & Section~\ref{sec:basics}, preceding Claim~\ref{cl:eq}. \\
$\textbf{X}_H(n,a)$ & A code that can correct a single substitution error. & Section~\ref{sec:basics}, preceding Claim~\ref{cl:eq}. \\
\hline
$\cB_{(T,\ell)}(\bfx)$ & The set of words obtained from $\bfx$ via $\ell$ adjacent transpositions.  & Section~\ref{sec:main}, preceding Example~\ref{ex:tdtderrors}. \\
$\cB_{(T,\ell),D}(\bfx)$ & The set of words obtained from $\bfx$ via $\ell$ adjacent transpositions and a single deletion. & Section~\ref{sec:main}, preceding Example~\ref{ex:tdtderrors}. \\
$\cC_{VT}(n,a,\ell)$ & \begin{tabular}{@{}c@{}}A VT-type code taken with modulus given by the parameter $\ell$. \\ 
The code $\cC_{VT}(n,a,b,\ell)$ comprises a subset of codewords in $\cC_{VT}(n,a,\ell)$ dictated by the parameter $b$. \end{tabular} & Section~\ref{sec:main}, following Lemma~\ref{lem:order}. \\
$\cD_{VT,n,\ell}$ & A decoder for $\cC_{VT}(n,a,\ell)$. &  Section~\ref{sec:main}, following Lemma~\ref{lem:order}. \\
$\cD_{VT,n,b,\ell}$ & A decoder for $\cC_{VT}(n,a,b,\ell)$. & Section~\ref{sec:main}, following Lemma~\ref{lem:similar2}.\\
$\cC_{(T,\ell) \land D}(n,a,b)$ & \begin{tabular}{@{}c@{}} A code which may correct a single deletion and up to $\ell$ adjacent transpositions. \\ $\cC_{(T,\ell) \land D}(n,a,b)$ is a subset of words in $\cC_{VT}(n,a,b,\ell)$. \end{tabular} & Section~\ref{sec:main}, before Theorem~\ref{th:general}. \\
$\textbf{Y}_{T \land D}(n,a_1, a_2)$ & A code used in the definition of $\cC_{T \land D}(n,a_1, a_2)$. & Section~\ref{sec:main}, following Corollary~\ref{cor:ellTD}. \\
$\cC_{T \land D}(n,a_1, a_2)$ & A code that may correct one adjacent transposition and one deletion. & Section~\ref{sec:main}, following Corollary~\ref{cor:ellTD}.\\
\hline
$\cB_{D,\leq b}(\bfx)$ & The set of words that may be obtained from $\bfx$ via a burst of consecutive deletions of length at most $b$. & Section~\ref{subsec:oddB}, Part 1. \\ 
$\cB_{D,b}(\bfx)$ & The set of words that may be obtained from $\bfx$ via a burst of consecutive deletions of length exactly $b$. & Section~\ref{subsec:oddB}, Part 1. \\ 
$\cC_{par}(n,b,\bfd)$ & A code used to determine the weight of a deleted substring. & Section~\ref{subsec:oddB}, Part 1. \\
$I(\bfy, \bfv, k_I)$ & A vector obtained by inserting $\bfv$ into $\bfy$ at position $k_I$. & Section~\ref{subsec:oddB}, preceding Claim~\ref{cl:vtBurstD}. \\
$D(\bfy,b,k_D)$ & A vector obtained by deleting $b$ consecutive bits from $\bfy$ starting at position $k_D$. & Section~\ref{subsec:oddB}, preceding Claim~\ref{cl:vtBurstD}. \\
$Bal(n,b)$ & A (balanced) set of words in which any sufficiently long substring has roughly half ones and half zeros. & Section~\ref{subsec:oddB}, preceding Claim~\ref{cl:CSB}. \\
$\cC^{odd}_b(n,a, \bfD)$ & \begin{tabular}{@{}c@{}}A code for determining the approximate location of a burst of deletions. \\ The code $\cC_{VT}(n,a,b,\ell)$ comprises a subset of words in $\cC_{VT}(n,a,\ell)$. \end{tabular} & Section \ref{subsec:oddB}, following Claim~\ref{cl:CSB}. \\
$SVT_{c,d}(n,M)$ & A code for determining the exact location of a deletion given an approximate location for the same. & Section \ref{subsec:oddB}, Part 3.\\
$\cC^{odd}_{b}(n,a, \bfC, \bfD)$ &  \begin{tabular}{@{}c@{}}A code which may correct a burst of deletions of odd length.  \\ The code $\cC^{odd}_{b}(n,a, \bfC, \bfD)$ is constructed using the codes $\cC^{odd}_b(n,a, \bfD)$ and $SVT_{c,d}(n,M)$. \end{tabular} & Section~\ref{subsec:oddB}, preceding Theorem~\ref{th:oddburst}. \\
$\cC_b(n,\bfa, \vec{\bfC}, \vec{\bfD})$ & \begin{tabular}{@{}c@{}}A code capable of correcting a burst of deletions of any length $\leq b$.   \\ $\cC_b(n,\bfa, \vec{\bfC}, \vec{\bfD})$  is constructed using the code $\cC^{odd}_{b}(n,a, \bfC, \bfD)$. \end{tabular} & Section~\ref{sec:general}, following Example~\ref{ex:general}.\\
\hline
$\cB_{BT,b}(\bfx)$ & The set of words obtained from $\bfx$ via one adjacent block transposition. & Section~\ref{sec:blockTransDel}, preceding Example~\ref{ex:extrans}. \\
$\cB_{BT \land D, b}(\bfx)$ & The set of words obtained from $\bfx$ via one adjacent block transposition and one block deletion. & Section~\ref{sec:blockTransDel}, following Example~\ref{ex:BTD}. \\
$T(\bfx, k_T)$ & The vector resulting from transposing the symbols at positions $k_T$ and $k_{T}+1$ in $\bfx$. & Section~\ref{sec:blockTransDel}, preceding Lemma~\ref{lem:svttdec}.\\
$\cC^{(1)}_{TD,b}(n,a, \bfC, \bfD)$ & A code for determining the approximate location of a block of deletions and adjacent transposition. & Section~\ref{sec:blockTransDel}, following Lemma~\ref{lem:svttdec}. \\
$\cC(n, m ; t_1, t_2)$ & A code for correcting special types of burst errors. & Section\ref{sec:blockTransDel}, following Definition~\ref{def:tpcodes}. \\
$\cC^{Odd,B}_{b}(n,a, \bfC, \bfD)$ & A code for correcting an odd-length block of deletions and adjacent block transposition. & Section~\ref{sec:blockTransDel}, following Lemma~\ref{lem:svttdec}. \\
$\cC_{TD,b}(n,\bfa, \vec{\bfC},\vec{\bfD})$ & A code for correcting one block of deletions and one adjacent block tranposition. & Section~\ref{sec:blockTransDel}, before Theorem~\ref{th:burstcode2}.
\end{tabular} 
\end{center}
\caption {Relevant Notation and Terminology.}
\end{table*}

Similar notation will be used for other types of editing errors, balls, distances and codes, with their meaning apparent from the context. Furthermore, for the convenience of the reader, relevant notation and terminology referred to throughout the paper is summarized in Table~I.

\section{Single Transposition or Deletion-Correcting Codes}\label{sec:basics}

We start by describing a general construction for single transposition \emph{or} deletion-correcting codes. 

We then show how to use this construction in order to devise codes with near-optimal redundancy.

Let $\cC_H(n,3)$ be a single-error-correcting code, and, as before, let $\cC_D(n)$ be a single-deletion-correcting code. We define a code $ \cC_{T \lor D}(n)$, which we show in Lemma~\ref{lem:t1} is capable of correcting one transposition (T) or ($\lor$) one deletion (D) as follows:
\begin{align}\label{cons:T-D codes}
 \cC_{T \lor D}(n)= \{ \bfx \in \mathbb{F}_2^n : \bfx\in \cC_{D}(n), \overline{\bfx} \in \cC_H(n,3) \}.
\end{align}

The code $\cC_{T \lor  D}(n)$ consists of codewords that belong to a single deletion error-correcting code and have integrals that belong to a single substitution error-correcting code. 
\begin{lemma}\label{lem:t1} 
The code $ \cC_{T \lor D}(n)$ described in (\ref{cons:T-D codes}) can correct a single adjacent transposition or a single deletion.

\end{lemma}
\begin{IEEEproof} 
We prove this claim by showing that for all $\bfx \in \cC_{T \lor D}(n)$, one can uniquely recover $\bfx$ from any $\bfz \in \cB_{T \lor D}(\bfx)$. 

Assume first that $\bfz \in \mathbb{F}_2^{n-1}$, so that $\bfz$ is the result of a single deletion occurring in $\bfx$. Since $ \bfx \in \cC_{D}(n)$, one may apply the decoder of the code $\cC_D(n)$ to successfully recover $\bfx \in \cC_{T \lor D}(n)$.

Assume that $\bfz \in \mathbb{F}_2^{n}$, so that $\bfz$ is the result of at most one single transposition in $\bfx$. We show that $d_H(\overline{\bfx}, \overline{\bfz}) \leq 1$. When this inequality holds, since $\overline{\bfx}$ belongs to a code with minimum Hamming distance $3$, the vector $\overline{\bfx}$ can be uniquely determined based on $\overline{\bfz}$. Note that since the mapping $\partial$ is injective, $d_H(\overline{\bfx}, \overline{\bfz}) =0$ if and only if ${\bfx} = {\bfz}$.

Let the transmitted word $\bfx$ be subjected to one adjacent transposition involving the $i$th and $(i+1)$th bits, so that $x_i \neq x_{i+1}$ and $\bfz= (x_1,\ldots,x_{i-1},x_{i+1},x_i,x_{i+2},\ldots, x_n)$. First, we compute the integral $\overline{\bfz}$ as
$$\overline{\bfz}= (z_1,z_2+z_1,z_3+z_2+z_1,\ldots,\sum_{j=1}^n z_j)=(\overline{z}_1, \ldots, \overline{z}_n).$$
Let $\overline{\bfx} = ( \overline{x}_1, \ldots, \overline{x}_n)$. Then, clearly $(\overline{x}_1, \ldots, \overline{x}_{i-1}) = (\overline{z}_1, \ldots, \overline{z}_{i-1})$. Furthermore, 
$$\overline{z}_i = \sum_{j=1}^{i-1} x_j + x_{i+1} = \sum_{j=1}^{i-1} x_j + (1 + x_{i}) = 1+\bar{x}_i,$$ 
and for any $k \geq i+1$, $\overline{z}_{k} = \sum_{j=1}^{i-1} x_j + x_{i+1} + x_i + \sum_{j=i+2}^{k} x_j = \overline{x}_{k}$, so that $d_H(\overline{\bfx}, \overline{\bfz}) = 1$ as desired.
\end{IEEEproof}

Observe that we did not explicitly state the choices of codes in (\ref{cons:T-D codes}). A natural choice would be a single substitution-correcting Hamming code, for which one requires that $n=2^m -1$ for some positive integer $m$, and the single deletion-correcting Varshamov-Tenengoltz (VT) code~\cite{levenshtein1966binary}, or some cosets of these codes. Since the cosets of the codes cover $\mathbb{F}_2^n$, one can see that there exists a code with redundancy at most $2\log (n+1)$. We show next how to improve this result by constructing \emph{one code} that may serve both as a single deletion-correcting codefor $\bfx$ and a single substitution-correcting code for $\overline{\bfx}$. The redundancy of this code is at most $\log\,n+\log6$.

Our choice of codes is as follows. Let $a$ be a non-negative integer such that $0\leq a\leq 6n-4$.  For the single deletion code, we use
\begin{align*}
\textbf{X}_D(n,a) =& \{ \bfx \in \mathbb{F}_2^n:  \sum_{i=1}^{n-1} i \, x_i + \\
&\ \ (2n-1) \, x_n \equiv \,a\, \bmod (6n-3) \}.
\end{align*}
For the code $\cC_H(n,3)$, we choose 
\begin{align*}
\textbf{X}_H(n,a) &= \{ \bfx \in \mathbb{F}_2^n:  \sum_{i=1}^{n-2} (2i+1) \, x_i + (2n-1) \, x_{n}  \notag \\
&+(3n-2) \, x_{n-1}  \equiv a \bmod (6n-3) \bigg\}. \notag
\end{align*}
\begin{claim}\label{cl:eq}
For any vector $\bfx\in\mathbb{F}_2^n$, if $\bfx' \in \textbf{X}_D(n,a)$ then $\bfx \in \textbf{X}_H(n,a)$ and thus if $\bfx\in \textbf{X}_D(n,a)$ then $\overline{\bfx} \in \textbf{X}_H(n,a)$.
\end{claim}
\begin{IEEEproof} Suppose that $\bfx' \in \textbf{X}_D(n,a)$. By definition,
$$\sum_{i=1}^{n-1} i \, x_i' + (2n-1) \, x_n' \equiv a \bmod 6n - 3.$$
Therefore, since $\bfx' = (x_1, x_1+x_2, x_2+x_3, \ldots, x_{n-1} + x_n )$, we have
\begin{align*}
x_1 + \sum_{i=2}^{n-1} i  \, (x_i + x_{i-1}) + (2n-1)  \, (x_{n-1} + x_n) &\equiv a \\
&\bmod 6n -3,
\end{align*}
which implies that $\bfx \in \textbf{X}_H(n,a)$. This proves the claim.
\end{IEEEproof}

According to Claim~\ref{cl:eq} and Lemma~\ref{lem:t1}, in order to show that the code $\cC_{T \lor D}(n)=\textbf{X}_D(n,a)$ is a single transposition or deletion-correcting code, we only have to show that the codes $\textbf{X}_D(n,a)$ and $\textbf{X}_H(n,a)$ have the desired error-correcting properties.

\begin{lemma} 
The code $\textbf{X}_{H}(n,a)$ is a single substitution error-correcting code. 
\end{lemma}

\begin{IEEEproof} Let $H=(3, 5, 7, \ldots,$ $2n-3, 3n-2, 2n-1)$ so that $\bfx \in \textbf{X}_{H}(n,a)$ if and only if $H \, \bfx^T \equiv a \bmod (6n-3)$. Assume on the contrary that $\textbf{X}_{H}(n,a)$ is not a single substitution error-correcting code. Then, there exist two different codewords $\bfx_1,\bfx_2 \in \textbf{X}_{H}(n,a)$ and two vectors $\bfe_j,\bfe_k$ such that $\bfx_1+\bfe_j = \bfx_2+\bfe_k$, where both $\bfe_j,\bfe_k$ have at most one non-zero entry of value either $1$ or $-1$. This would imply
\begin{align*}
H \, (\bfx_1 + \bfe_j)^T & \equiv H \, (\bfx_2 + \bfe_k)^T \bmod (6n -3), \text{ and}\\
H \, \bfe_j^T &\ \equiv \ H \, \bfe_k^T \bmod (6n -3),
\end{align*}
which holds if and only if $\bfe_j=\bfe_k$. Therefore, we must have $\bfx_1=\bfx_2$, a contradiction.
\end{IEEEproof}

\begin{lemma} 
The code $\textbf{X}_D(n,a)$ can correct a single deletion.
\end{lemma}
\begin{IEEEproof} By definition, if $\bfx \in \textbf{X}_D(n,a)$, we may write
$$ H \, \bfx^T \equiv a \bmod 6n - 3, $$
where $H = (1,2,3,\ldots, n-1, 2n-1)$. The result follows by observing that $(1,2,3\ldots, n-1, 2n-1)$ is a Helberg sequence as defined in Definition III.2 from~\cite{hagiwara2016short}. Thus, according to Theorem III.4 of the same paper, the code $\textbf{X}_D(n,a)$ can correct a single deletion.
\end{IEEEproof}

The following corollary summarizes the main result of this section.
\begin{corollary} 
There exists a single transposition or deletion-correcting code whose redundancy is at most $\log(6n-3)$ bits. 
\end{corollary}
\begin{IEEEproof}

Using the pigeon-hole principle considered in~\cite{sloane2002single}, one may easily show that $ | \textbf{X}_{H}(n,a)| = | \cC_{T \lor D}(n,a)| \geq \frac{2^n}{6n-3},$ since $\cC_{T \lor D}(a,n)$ partitions the ambient space $\mathbb{F}_2^n$ into $6n-3$ codes, one of which has to have a size at least as large as the right-hand side of the inequality.
\end{IEEEproof}

Note that every single transposition or deletion-correcting code is also a single deletion error-correcting code. Hence, a lower bound on the redundancy of the latter code is $\log\,n$~\cite{6497614}, so that the difference between the redundancy of our deletion/adjacent transposition codes and the redundancy of a optimal single deletion code is at most $\log 6$ bits. We also note that improving the lower bound on a single transposition or deletion-correcting code is left as an open problem.

\section{Codes Correcting Deletions and Adjacent Transpositions} \label{sec:main}

We now turn our attention to the significantly more challenging task of constructing codes that can correct both deletions and adjacent transpositions simultaneously. Our main result is a construction of a code capable of correcting a single deletion along with multiple adjacent transpositions. At the end of this section, we present an improved construction for the special case of a single deletion and a single transposition.

We start by introducing some useful notation. Let $\cB_{(T,\ell)}(\bfx)$ denote the set of vectors that may be obtained by applying at most $\ell$ adjacent transpositions (T) to $\bfx$. Hence,
$$\cB_{(T,\ell)}(\bfx)= \underbrace{\cB_{(T,1)}( \ldots ( \cB_{(T,1)}(}_{\text{$\ell$ times}} \bfx ) ) \ldots ).$$

Let $\cB_{(T,\ell),D}(\bfx)$ denote the set of vectors that may be obtained from $\bfx$ by at most $\ell$ adjacent transpositions followed by at most one single deletion. As before, let $\cB_D(\bfx)$ be the set of words that may be obtained by introducing at most one deletion into $\bfx$. With a slight abuse of notation, we use the same symbol $\cB$ independently on the the argument of the set being a word or a collection of words. In the latter case, the set $\cB$ equals the union of the corresponding sets of individual words in the argument. The next example illustrates the relevant notation.

\begin{example}\label{ex:tdtderrors} Suppose that $\bfx=(0,0,1,1,0)$. Then, 
$$\cB_{(T,1)}(\bfx) = \{ (0,0,1,1,0),(0,1,0,1,0),(0,0,1,0,1) \},$$
$$\cB_{D}(\bfx)= \{ (0,0,1,1,0),(0,1,1,0),(0,0,1,0),(0,0,1,1) \},$$ 
$$\hspace{-2.0ex}\cB_{(T,1),D}(\bfx) = \{ (0,0,1,1,0),(0,1,1,0),(0,0,1,0),(0,0,1,1),(1,0,1,0)$$
$$(0,1,0,1),(0,1,0,0),(0,0,0,1), (0,1,0,1,0), (0,0,1,0,1) \}.$$
\end{example}

\begin{lemma}\label{lem:order} For any $\bfx \in \mathbb{F}_2^n$, 
$$\cB_{(T,\ell),D}(\bfx) = \cB_D(\cB_{(T,\ell)}(\bfx)) = \cB_{(T,\ell)}(\cB_D(\bfx)).$$ 
\end{lemma}
\begin{IEEEproof} 
The proof is by induction on $\ell$. For the base case $\ell=1$, we show that $\cB_D(\cB_{(T,1)}(\bfx)) = \cB_{(T,1)}(\cB_D(\bfx))$ by demonstrating that if $\bfy \in \cB_{(T,1)}(\cB_{D}(\bfx))$, then $\bfy \in \cB_D(\cB_{(T,1)}(\bfx))$. Furthermore if $\bfy \in \cB_D(\cB_{(T,1)}(\bfx))$, then $\bfy \in \cB_{(T,1)}(\cB_{D}(\bfx))$.

Suppose that $\bfy^{(d)} = (x_1, \ldots, x_{i-1}, x_{i+1}, \ldots x_n)$ is the result of deleting the symbol at position $i$, 
where $i \in [n]$. Also, assume that $\bfy=\bfy^{(d,t)}$ is obtained from $\bfy^{(d)}$ by transposing the symbol in position $j$ with the symbol in position $j+1$ in $\bfy^{(d)},$ where $j \in [n-2]$. One needs to consider two different scenarios: 1) $j \in [n-2] \setminus (i-1)$; and 2) $j = i-1$. 

First, we show that if $j \in [n-2] \setminus (i-1)$, then $\bfy \in \cB_{D}(\cB_{(T,1)}(\bfx))$. To see why this claim holds, note that if $j < i-1$ then $\bfy$ may be generated by first transposing the symbols in positions $j, j+1$ in $\bfx$ to obtain $\bfy^{(t)}$ and then deleting the symbol in position $i$. Otherwise, if $j \geq i$, one may first transpose the symbols in positions $j+1, j+2$, and then delete the symbol in position $i$. Suppose now that $j=i-1$. Then $x_{i-1} \neq x_{i+1}$ and so $x_i$ equals either $x_{i-1}$ or $x_{i+1}$. Suppose that $x_i = x_{i-1}$. Then $\bfy$ may be generated by first transposing $x_i$ and $x_{i+1}$, and then deleting the symbol in position $i-1$. Otherwise, if $x_i = x_{i+1}$, $\bfy$ may be obtained by first transposing $x_{i-1}$ and $x_i$ and then deleting the symbol in position $i+1$.

Using a similar argument, it can be shown that if $\bfy \in \cB_{D}(\cB_{(T,1)}(\bfx))$, then $\bfy \in \cB_{(T,1)}(\cB_{D}(\bfx))$. 
This establishes the base case $\cB_D(\cB_{(T,1)}(\bfx)) = \cB_{(T,1)}(\cB_D(\bfx))$.

We now prove the inductive step. 

Suppose that $\cB_D(\cB_{(T,\ell)}(\bfx)) = \cB_{(T,\ell)}(\cB_D(\bfx))$ holds for all $\ell < L$. We show that $\cB_D(\cB_{(T,L)}(\bfx)) = \cB_{(T,L)}(\cB_D(\bfx))$ holds as well. This may be seen from the following chain of equalities:
\begin{align*}
\cB_D(\cB_{(T,L)}(\bfx)) &= \cB_D(\cB_{(T,L-1)}(\cB_{(T,1)}(\bfx))) \\
&=\cB_{(T,L-1)}(\cB_D(\cB_{(T,1)}(\bfx))) \\
&=\cB_{(T,L-1)}(\cB_{(T,1)}(\cB_D(\bfx))) \\
&=\cB_{(T,L)}(\cB_D(\bfx)),
\end{align*}
where the second line follows from the inductive hypothesis, which is applied to each vector in the set, and where the third line is a result of the previous result which showed that $\cB_D(\cB_{(T,1)}(\bfx)) = \cB_{(T,1)}(\cB_D(\bfx))$.
\end{IEEEproof}

As a consequence of the previous lemma, we may henceforth assume that the deletion always occurs after the adjacent transposition(s). We then say that a code $\cC$ can correct $\ell$ adjacent transpositions and a single deletion, and refer to it as a \emph{$\ell$-TD code} if for any two different codewords $\bfu,\bfv\in \cC$, $\cB_{(T,\ell),D}(\bfu)\cap\cB_{(T,\ell),D}(\bfv) =\emptyset$. 
Our code construction and the ideas behind the coding approach are best explained through the decoding procedure. 

Suppose that the code $\cC_{T \land D}(n,\ell)$ is an $\ell$-TD code, which is a subset of codewords of a single deletion-correcting code. Assume also that  $\bfx \in \cC_{T \land D}(n,\ell)$ was transmitted and that the vector $\bfy$ was received, where $\bfy$ is the result of at most $\ell$ transpositions followed by at most one single deletion in $\bfx$. The simplest idea to pursue is to try to correct the single deletion by naively applying the decoder for the chosen constituent single-deletion code. Clearly, such a decoder may produce an erroneous result due to the presence of the adjacent transposition errors. It is therefore important to construct the code $\cC_{T \land D}(n,\ell)$ in such a way that the result of the ``mismatched'' deletion correction $\widehat{\bfx}$, obtained from $\bfy$, is easy to characterize and contains only a limited number of errors that may be corrected to recover $\bfx \in \cC_{T \land D}(n,\ell)$ from $\widehat{\bfx}$. To this end, define the following code:
\begin{align*}
\cC_{VT}(n,a,\ell) = \{ &\bfx \in \mathbb{F}_2^n : \sum_{i=1}^n i  x_i \equiv a \bmod (n+2\ell+1) \}. \nonumber
\end{align*}

Since the code is a VT code, the decoder $\cD_{VT,n,\ell}$ for $\cC_{VT}(n,a,\ell)$ can correct a single deletion occurring in any codeword in $\cC_{VT}(n,a,\ell)$~\cite{sloane2002single}. Note that the standard definition of a single deletion-correcting code entails setting $\sum_{i=1}^n i \, x_i$ to be equal to some $a$ modulo $n+1$ \cite{sloane2002single}. Our construction fixes $\sum_{i=1}^n i \, x_i$ to $a$ modulo $n+2\ell+1$ instead. As we demonstrate in Claim~\ref{cl:trans}, this change is needed due to the fact that adjacent transpositions may change the value of the syndrome $a$ by at most $\pm \ell$.  

As before, and for the special case of VT codes, assume that $\widehat{\bfx}$ is the result of VT decoding the vector $\bfy$ where $\bfy\in\cB_{(T,\ell),D}(\bfx)$. Our first aim is to characterize the difference between $\widehat{\bfx}$ and $\bfx$, and for this purpose we use an intermediary word $\bfy^{(\ell)}$ that is generated from at most $\ell$ adjacent transpositions in $\bfx$, i.e., a word such that $\bfy \in \cB_D(\bfy^{(\ell)})$. 

More precisely, we demonstrate that if both $\bfx, \bfy^{(\ell)} \in \cC_{VT}(n,a,\ell)$, then the decoder outputs $\cD_{VT,n,\ell}(a,\bfy)$ and $\cD_{VT,n,\ell}(a,\bfy^{(\ell)})$ will differ only in the transpositions that actually occurred in $\bfx$. On the other hand, if $\bfx, \bfy^{(\ell)}$ belong to two different VT codes (i.e. they have different values of the VT syndrome parameter $a$), then $\bfx$ and $\widehat{\bfx}$ differ by at most $2\ell$ adjacent transpositions. The following simple claim is a consequence of the fact that an adjacent transposition changes the VT syndrome by at most one.
\begin{claim}\label{cl:trans} Suppose that $\bfy^{(\ell)}=(y^{(\ell)}_1, \ldots, y^{(\ell)}_n) \in \cB_{(T,\ell)}(\bfx)$ where $\bfx \in \mathbb{F}_2^n$. Then, one has
$| \sum_{i=1}^n i \, x_i - \sum_{i=1}^n i \, y^{(\ell)}_i | \leq \ell.$
\end{claim}
\begin{IEEEproof} The proof is by induction on $\ell$. For the base case suppose $\bfy^{(1)} \in \cB_{(T,1)}(\bfx)$. The result clearly holds if $\bfy^{(1)} = \bfx$ and so assume $\bfy^{(1)}$ is the result of transposing the symbols in positions $j$ and $j+1$ in $\bfx$. Then 
\begin{align*}
| \sum_{i=1}^n i \, x_i &- \sum_{i=1}^n i \, y^{(\ell)}_i |\\
&= \Big | i \, x_i + (i+1) \, x_{i+1} - \Big (i \, x_{i+1} + (i+1) \, x_i \Big ) \Big |\\
&=|x_{i+1} - x_i| =1,
\end{align*}
since $x_i \neq x_{i+1}$. For the inductive step, suppose that the result holds for all $\ell < L$ and consider the case $\ell=L$. Let $\bfy^{(L)} \in \cB_{(T,L)}$ and let $\bfy^{(L-1)} \in \cB_{(T,L-1)}$ be such that $\bfy^{(L)}$ and $\bfy^{(L-1)}$ differ by at most one single adjacent transposition. Then, 
\begin{align*}
&|\sum_{i=1}^n i \, x_i - \sum_{i=1}^n i \, y^{(L)}_i | \\
=&| \sum_{i=1}^n i \, x_i - \sum_{i=1}^n i \, y^{(L-1)}_i + \sum_{i=1}^n i \, y^{(L-1)}_i - \sum_{i=1}^n i \, y^{(L)}_i |\\
\leq &| \sum_{i=1}^n i \, x_i - \sum_{i=1}^n i \, y^{(L-1)}_i | + | \sum_{i=1}^n i \, y^{(L-1)}_i - \sum_{i=1}^n i \, y^{(L)}_i | \\
\leq &L -1 + 1=L.
\end{align*}
\end{IEEEproof}

As a consequence of the previous claim, if $\bfx \in\cC_{VT}(n,a,\ell)$ and $\bfy^{(\ell)} \in \cB_{(T,\ell)}(\bfx)$, then $\bfy^{(\ell)} \in \cC_{VT}(n,\hat{a},\ell)$ for some $\hat{a}$, where 
$ | a - \hat{a} | \leq \ell. $ The next lemma summarizes the previous discussion.

\begin{lemma}\label{lem:equiv} Suppose that $\bfy^{(\ell)} \in \cB_{(T,\ell)}(\bfx),$ where $\bfx \in \cC_{VT}(n,a,\ell)$, and let $\bfy\in\cB_{D}(\bfy^{(\ell)})$. Then, $\cD_{VT,n,\ell}(\hat{a},\bfy)=\bfy^{(\ell)}$ for some $\hat{a}$ such that $|a - \hat{a}| \leq \ell$. 
\end{lemma}

\begin{example} Suppose that $\bfx=(0,1,1,0,0,1,0,0,0,0,1,0) \in \cC_{VT}(12, 3, 3)$ was transmitted and that the vector 
$\bfy = (0,1,1,0,$ $0,1,0,0,$ $1,0,0) $ was received after at most three adjacent transpositions and a single deletion.
For $\bfy^{(3)} =  (0,1,1,0,$ $0,1,0,0,$ $0,1,0,0)$ (where $\bfy \in \cB_D(\bfy^{(3)})$, we have $\sum_{i=1}^{n-1} i \cdot y_i \equiv 2 \bmod 19$. Thus, since $a=3$ and $\hat{a}=2$, we get that $|a - \hat{a}| \leq 1 \leq \ell = 3$ as desired.

Note that if we use the decoder $\cD_{VT,12,3}$ we arrive at $\widehat{\bfx}=\cD_{VT,12,3}(3,\bfy) = (0,1,1,0,0,0,1,0,0,1,0,0)$. Hence, we have
$\widehat{\bfx}=(0,1,1,0,0,0,1,0,0,1,0,0),$ and
$\bfx=(0,1,1,0,0,1,0,0,0,0,1,0).$
\end{example}

We characterize next the difference between $\cD_{VT,n,\ell}(a, \bfy)$ and $\cD_{VT,n,\ell}(\hat{a}, \bfy)$ for the case that $|a - \hat{a}| \leq \ell$, as the value $\hat{a}$ is not known beforehand. 

Our main result may be intuitively described as follows: Suppose that $\bfy \in \cB_D(\bfx),$ where $\bfx \in \cC_{VT}(n,a,\ell)$ and where $\bfy$ is obtained by deleting the $k$th bit, $x_k$, from $\bfx$. Also, assume that the value of $x_k$ is known to the decoder and that $\widehat{\bfx}=\cD_{VT,n,\ell}(a+v,\bfy)$, for some offset $v$, is obtained by inserting the bit $x_k $ into $\bfy$ at some position determined by the decoder. 
Then, if $x_k=0$, we may obtain $\bfx$ from $\widehat{\bfx}$ by sliding the inserted bit to the left/right using a series of adjacent transposition operations past at most $v$ ones. Otherwise, if $x_k=1$, then we can obtain $\bfx$ from $\widehat{\bfx}$ by sliding the inserted bit to the left/right past at most $v$ zeros. The next lemma rigorously summarizes this observation.

\begin{lemma}\label{lem:similar2} Suppose that $\bfy$ is the result of a single deletion occurring in $\bfx \in \cC_{VT}(n,a,\ell)$ at position $k$. Given $k$, let $v_L=|\{ j \in [n] : j < k, x_j=1 \}|$ and $v_R=|\{ j \in [n] : j > k, x_j=1 \}|$. Then, 
\begin{enumerate}
\item If $x_k=0$, then for all $v \in \{-v_R, -v_R+1, \ldots, v_L\}$, one may obtain $\cD_{VT,n,\ell}(a+v,\bfy)$ by inserting the symbol $0$ into $\bfy$ immediately after the $(v_L-v)$-th one. 
\item If $x_k=1$, then for all $v \in \{-(k-1)+v_L, -k+v_L+2,$ $\ldots, (n-k)-v_R \}$, one may obtain $\cD_{VT,n,\ell}( a + v,\bfy)$ by inserting the symbol $1$ into $\bfy$ immediately after the $(v+k-v_L-1)$-th zero.
\end{enumerate}
\end{lemma}

\begin{example} Suppose that $\bfx=(0,1,1,0,0,1,0,\textcolor{black}{0},0,0,1,0) \in \cC_{VT}(12, 3, 3)$, and that $\hat{\bfx} = \cD_{VT,n,\ell}(3,\bfy),$
was obtained by VT decoding $\bfy = (0,1,1,0,1,0,\textcolor{black}{0},0,0,1,0)$. For $v=2$, one has 
$\cD_{VT,n,\ell}(5,\bfy)=(0,0,1,1,0,1,0,0,0,0,1,0),$
whereas for $v=-1$, one has 
$\cD_{VT,n,\ell}(2,\bfy)=({0},1,1,0,1,0,0,0,0,0,1,0).$

Next, suppose that $\bfy = (0,1,1,0,0,0,0,0,0,1,0)$, where $\bfy$ is the result of deleting the third $1$ at position $k=6$ from $\bfx=(0,1,1,0, \textcolor{black}{0},$ $1,0,\textcolor{black}{0},$ $0,0,1,0)$. In this case, choosing $v=3$ gives $ \cD_{VT,n,\ell}(6,\bfy)=(0,1,1,0,0,0,0,0,1,0,1,0)$, while $v=-2$ gives 
$ \cD_{VT,n,\ell}(1,\bfy)=(0,1,1,1,0,0,0,0,0,0,1,0).$
\end{example}

{ {\textit{Proof of Lemma~8:}} Suppose first that $\bfy$ is the result of deleting a zero from $\bfx \in \cC_{VT}(n,a,\ell)$. Let $a' \equiv a - \sum_{i=1}^{n-1} i \, y_i \bmod (n+ 2\ell+1)$. The decoder $\cD_{VT,n,\ell}$ for $\cC_{VT}(n,a,\ell)$ produces the vector $\widehat{\bfx} \in \cC_{VT}(n,a,\ell)$ by inserting a zero into the first position $k'$ that has $a'$ ones to the right of it. If $x_k=0$, then clearly $a' = v_R$, and the decoder correctly outputs $\bfx$ so that $\widehat{\bfx} = \bfx$. If the decoder $\cD_{VT,n,\ell}$ for $\cC_{VT}(n,a+v,\ell)$ were applied to $\bfy$ instead, one would have
$$a'' \equiv a+v - \sum_{i=1}^{n-1} i \, y_i \bmod (n+2\ell+1) \equiv a' + v \bmod (n+2\ell+1). $$
Hence, the decoder $\cD_{VT,n,\ell}$ for $\cC_{VT}(n,a+v,\ell)$ would insert a zero in the vector $\bfy$ at the first position $k''$ that has $a' + v$ ones to the right of it. The claim follows by observing that the position immediately following $a'+v$ ones is in the same run as the position in $\bfy$ preceding $(v_L-v)$ ones.

Suppose next that $\bfy$ is the result of deleting a one from $\bfx \in \cC_{VT}(n,a,\ell)$. Let $a' \equiv a - \sum_{i=1}^{n-1} i \, y_i \bmod (n+2\ell+1)$. The decoder $\cD_{VT,n,\ell}$ for $\cC_{VT}(n,a,\ell)$ produces the vector $\widehat{\bfx} \in \cC_{VT}(n,a,\ell)$ by inserting a one into the first position $k'$ with $a'-k'$ ones its right. If $x_k=1$, then clearly $k' = k$ and the decoder correctly outputs $\bfx$, so that $\widehat{\bfx} = \bfx$. Note that position $k$ appears before $v_R = a'-k$ ones and after $k-1-v_L$ zeros (i.e., position $k$ has $a'-k$ ones on its right and $k-1-v_L$ zeros to its left). Furthermore, the total number of ones in $\bfx$ is $v_L + v_R + 1 = v_L + a' - k + 1$, which implies that
\begin{align}\label{eq:totalones}
v_L + v_R =v_L + a' - k. 
\end{align}

If the decoder $\cD_{VT,n,\ell}$ for $\cC_{VT}(n,a+v,\ell)$ were applied to $\bfy$ instead, then one would have $a''  \equiv a' + v \bmod (n+2\ell+1)$ as before. The decoder $\cD_{VT,n,\ell}$ for $\cC_{VT}(n,a+v,\ell)$ would insert a one into the vector $\bfy$ at the first position $k''$ preceeding $a' + v - k''$ ones (or with $a'+v-k''$ ones to its right). This produces a vector $\widehat{\bfx}$. Given (\ref{eq:totalones}), since the total number of ones in $\bfx$ is $v_L + v_R + 1$, we know that the number of ones preceding position $k''$ (i.e., to its left) is 
$$ v_L + a' - k - (a' + v - k'') = v_L - k - v + k''. $$
Thus, the number of zeros preceeding $k''$ (or to its left) is 
$$ (k'' - 1) - (v_L - k - v + k'') = k + v - v_L - 1, $$
which proves the claim of the lemma.
$\;\;\;\;\;\;\;\;\;\;\;\;\;\;\;\;\;\;\;\;\;\;\;\;\;\;\;\;\;\;\;\;\;\;\;\;\;\;\;\;\;\;\;\;\;\;\;\;\;\;\;\;\;\;\;\;\;\;\;\;\;\;\;\;\;\;\;\;\;\;\;\;\;\;\;\;\;\blacksquare$} \\

The previous lemma motivates the introduction of a modification of VT codes, which will be used as a constituent component in a construction of codes capable of correcting a deletion and multiple adjacent transpositions. This modified code structure also leads to a straightforward decoding procedure of the underlying codes. 
The code may be defined as follows:
\begin{align}\label{eq:vtphi}
\cC_{VT}(n,a,b,\ell) = \{ &\bfx \in \mathbb{F}_2^n : \\
&\sum_{i=1}^n i \, x_i \equiv a \bmod (n+2\ell+1), \nonumber \\
&\sum_{i=1}^{n} x_i \equiv b \bmod 2 \}. \nonumber
\end{align}

The code $\cC_{VT}(n,a,b,\ell)$ allows one to first determine the value of the deleted bit using the second parity constraint and then subsequently determine the location of the deleted bit using the VT-type constraint.

The decoder for  $\cC_{VT}(n,a,b,\ell)$, denoted by $\cD_{VT,n,b,\ell}$, operates as follows. Suppose that $\bfx \in \cC_{VT}(n,a,b,\ell)$ is transmitted and that $\bfy \in \cB_{(T,\ell),D}(\bfx)$ is received. Suppose that $n_1$ denotes the number of ones in $\bfy$. Then, for $a \in \mathbb{Z}_{n+2\ell+1}$ and $b \in \mathbb{F}_2$,  $\cD_{VT,n,b,\ell}(a,\bfy)$ executes the following steps:
\begin{enumerate}
\item Set $x \equiv \sum_{i=1}^{n-1} {y_i} + b \bmod 2$.
\item Compute $a' \equiv a- \sum_{i=1}^{n-1} i \, y_i \bmod (n+2\ell+1).$
\item If $x=0$ and $a' \in \{0,1,\ldots, n_1 \}$, insert a zero into the first position in $\bfy$ that has $a'$ ones on its right. \\
If $a' \in \{n_1+1,n_1+2,\ldots,n_1+\ell \}$, insert a zero in the first position in $\bfy$. \\
If $a' \in \{ {n}+\ell+1, {n}+\ell+2, \ldots, {n}+2\ell\}$, insert a zero in the last position of $\bfy$.
\item If $x=1$ and $a' \in \{n_1+1,n_1+2,\ldots,n\}$, insert a one in the first position $k$ of $\bfy$ that has $a'-k$ ones to its right. \\
Otherwise, if $a' \in \{n+1,n+2,\ldots,n+\ell\}$, insert a one in the last position of $\bfy$. \\
If $a' \in \{ {n_1}-\ell+1,n_1-\ell+2,\ldots,n_1\}$, insert a one in the first position of $\bfy$.
\end{enumerate}

Note that the VT decoder discussed so far aims to correct a single deletion only, but potentially in a mismatched fashion as additional adjacent transposition errors may have been incurred during deletion correction. 
The output of the deletion-correcting decoder has to be fed into the input of a transposition error-correcting code, and we describe how this subsequent decoding is accomplished after providing an 
illustration of the VT decoding process.

\begin{example}  Suppose that $\bfx=(0,1,1,0,0,1,0,\textcolor{black}{0},0,0,1,0) \in \cC_{VT}(12, 3,0,3)$, and that
$\bfy = (0,1,1,0,1,0,\textcolor{black}{0},0,1,0,0)$
is the received word, which is the result of a single deletion and a single transposition. We first apply the decoder $\cD_{VT,12,0,3}$ to $\bfy$. In the first step of the procedure, we conclude that the deleted bit has value $x=0$. In the second step of decoding, we compute $a' = 3$. Since $0 \leq a' \leq 4$, we have $\widehat{\bfx}= (0,1,0,1,0,1,0,\textcolor{black}{0},0,1,0,0)$. Note that
$\widehat{\bfx}= (0,1,0,1,0,1,0,\textcolor{black}{0},0,1,0,0),$ and
$\bfx= (0,1,1,0, \textcolor{black}{0},1,0,\textcolor{black}{0},0,0,1,0),$
differ in \emph{two} adjacent transpositions.
\end{example}

The previous example illustrates that $\bfx$ and $\widehat{\bfx}$ differ in a limited number of transpositions which depends on the original number of transposition errors. In particular, for the given example, the two vectors differ in two adjacent transpositions as $\widehat{\bfx}$ is the result of a single deletion and a single transposition in $\bfy$. The next lemma gives a more precise characterization of the ``distance'' between $\bfx$ and  $\widehat{\bfx}$.

\begin{lemma}\label{lem:miserror}  Suppose that {$\bfy^{(\ell)} \in \cB_{(T,\ell)}(\bfx)$} where $\bfx \in  \cC_{VT}(n,a,b,\ell)$ and where $\bfy\in\cB_D(\bfy^{(\ell)})$. Let $\widehat{\bfx} = \cD_{VT,n,b,\ell}(a,\bfy)$. Then the following statements are true:
\begin{enumerate}
\item If $\widehat{\bfx}$ is the result of inserting a zero in $\bfy$ in a position with $v^{(1)}_R$ ones to the right of the inserted bit, then $\bfy^{(\ell)}$ can be obtained from $\bfy$ by inserting a zero in $\bfy$ in the first position that precedes $j$ ones where $j \in \{ v^{(1)}_R-\ell, v^{(1)}_R - \ell+1, \ldots, v^{(1)}_R + \ell \}$.
\item If $\widehat{\bfx}$ is the result of inserting a one in $\bfy$ in position $k$ with $v^{(0)}_R$ zeros to the right of the inserted bit, then $\bfy^{(\ell)}$ can be obtained from $\bfy$ by inserting a one in $\bfy$ in the first position that precedes $j$ zeros where $j \in \{ v^{(0)}_R-\ell, v^{(0)}_R - \ell+1, \ldots, v^{(0)}_R + \ell \}$.
\end{enumerate}
\end{lemma}
\begin{IEEEproof} Suppose that $\widehat{\bfx} = \cD_{VT,n,b,\ell}(a,\bfy)$ is the result of inserting a zero into $\bfy$. According to Claim~\ref{cl:trans}, $\bfy^{(\ell)} \in \cC_{VT}(n, a + v, b, \ell)$ for some $v$, where $|v| \leq \ell$. Suppose next that $\bfy$ is the result of deleting a zero from $\bfy^{(\ell)}$ at position $k'$, where position $k'$ precedes $\tilde{v}^{(1)}_{R}$ ones in $\bfy^{(\ell)}$,  and position $k'$ follows $\tilde{v}^{(1)}_L$ ones. Clearly, $\bfy^{(\ell)} = \cD_{VT,n,b,\ell}(a + v,\bfy)$. According to Lemma~\ref{lem:similar2}, $\widehat{\bfx} = \cD_{VT,n,b,\ell}( (a+v) - v,\bfy)$ is obtained from $\bfy$ by inserting a zero into the first position with $\tilde{v}^{(1)}_L + v$ ones to its left and $\tilde{v}^{(1)}_R - v$ to its right, which proves the first statement in the lemma.

Suppose next that $\widehat{\bfx} = \cD_{VT,n,b,\ell}(a,\bfy)$ is the result of inserting a one into $\bfy$. Based on the same reasoning as the one used in the first part of the proof, we have $\bfy^{(\ell)} \in \cC_{VT}(n, a + v, b, \ell)$ for some $v$, where $|v| \leq \ell$. Suppose $\bfy$ is the result of deleting a one from $\bfy^{(\ell)}$ at position $k'$, where $k'$ is such that there are $\tilde{v}^{(1)}_{R}$ ones to the right of this position, and $\tilde{v}^{(1)}_L$ ones to the left of this position. Furthermore, we assume there are $\tilde{v}^{(0)}_{R}$ zeros to the right of position $k'$, and $\tilde{v}^{(0)}_L$ zeros to the left of position $k'$. Then, $\bfy^{(\ell)} = \cD_{VT,n,b,\ell}(a + v,\bfy)$. According to Lemma~\ref{lem:similar2}, $\widehat{\bfx} = \cD_{VT,n,b,\ell}( (a+v) - v,\bfy)$ is obtained from $\bfy$ by inserting a one into $\bfy$ after the $(k'-\tilde{v}^{(1)}_L-1-v)$-th zero. Equivalently, we can obtain $\widehat{\bfx}$ by inserting a one into $\bfy$ in the first position with $\tilde{v}_L^{(0)} - v$ zeros to its left and $\tilde{v}_R^{(0)} + v$ zeros to its right, since $\tilde{v}_L^{(0)} = (\textcolor{black}{k'}-1)- \tilde{v}_L^{(1)}$.
\end{IEEEproof}
The following corollary summarizes one of the main results of this section.

\begin{corollary}\label{cor:deltrans11} Suppose that $\bfy \in \cB_{(T,\ell),D}(\bfx)$ where $\bfx \in \cC_{VT}(n,a,b,\ell)$ and let $\widehat{\bfx}=\cD_{VT,n,b,\ell}(a,\bfy)$. Then $\bfx \in \cB_{(T,2\ell)}(\widehat{\bfx})$.
\end{corollary}
Consequently, the mismatched VT decoder increases the number of adjacent transposition errors by at most a factor of two.

Based on the results on mismatched VT decoding and Corollary~\ref{cor:deltrans11}, we are now ready to define a family of codes capable of correcting a single deletion and multiple adjacent transposition errors. Recall that given a binary word {$\bfx$}, its derivative $\partial(\bfx)=\bfx'$ is defined as $\bfx' = (x_1,x_2+x_1,x_3+x_2,\ldots, x_n+x_{n-1})$ and its inverse (integral) as $\partial^{-1}(\bfx)=\overline{\bfx} = (x_1, x_1 + x_2, \ldots, \sum_{i=1}^n x_i)$.
We claim that the code $\cC_{(T,\ell) \land D} \subseteq \mathbb{F}_2^n$
\begin{align}\label{eq:generaltrans}
\cC_{(T,\ell) \land D}(n,a,b)= \{ \bfx \in \mathbb{F}_2^n  :\ & \overline{\bfx} \in \cC_H(n,4\ell+1), \notag \\
&\bfx \in \cC_{VT}(n,a,b,\ell) \}
\end{align}
is an $\ell$-TD code (i.e., a code capable of correcting $\ell$ adjacent transpositions $(T,\ell)$ \emph{and} ($\land$) one deletion (D)). This result intuitively follows from the fact that the coupling of a VT-type constraint and a substitution error-correcting code with sufficiently large distance can handle a single deletion along with a number of adjacent transpositions, akin to what was established in the previous sections for the case of a single adjacent transposition.
\begin{theorem}\label{th:general} 
The code $\cC_{(T,\ell) \land D}(n,a,b)$ is an $\ell$-TD code.
\end{theorem}
\begin{IEEEproof} Suppose that $\bfy \in  \cB_{(T,\ell),D}(\bfx)$. We show how to recover $\bfx$ from $\bfy$. First, we determine $\widehat{\bfx} = \cD_{VT,n,b,\ell}(a,\bfy)$. From Corollary~\ref{cor:deltrans11}, we have that $\bfx \in \cB_{(T,2\ell)}(\widehat{\bfx})$. 
Since $\bfx \in \cB_{(T,2\ell)}(\widehat{\bfx})$, we have $d_H(\partial^{-1}(\widehat{\bfx}), \overline{\bfx}) \leq 2\ell$. Because the minimum distance of the code $\overline{\cC}_{(T,\ell) \land D}(n,a,b)$ is $4\ell + 1$, we can uniquely recover $\bfx$ from $\partial^{-1}(\widehat{\bfx})$.
\end{IEEEproof}

The following bound follows by noting the existence of binary codes of length $n$ and minimum distance $4\ell + 1$ which have $2\ell \log n$ bits of redundancy (see \cite{Roth}, Problem 8.12).
\begin{corollary}\label{cor:ellTD} There exists an $\ell$-TD code which redundancy at most $2\ell\log\,n + \log (n+2\ell+1)$ bits.
\end{corollary}
Next, we improve upon this result for the case when $\ell=1$. 

Let $a_1, a_2 \in \mathbb{Z}_{n+2L+1}$. Define $\textbf{Y}_{T \land D}(n,a_1,a_2) \subseteq \mathbb{F}_2^n$ according to
\begin{align*}
\textbf{Y}_{T \land D}(n,a_1, a_2) = &\{\bfx \ : x_n = 0, \\
&\sum_{i=1}^{n-1} (2i+1) x_i \equiv a_1 \bmod (n + 2L + 1), \\
&\sum_{i=1}^{n-1} (2i+1)^3 \, x_i   \equiv a_2 \bmod (n+2L+1) \},
\end{align*}
where $L \geq 1$ is chosen so that $n + 2L + 1$ is a prime number greater than $2n-1$. 

Let $$\cC_{T \land D}(n,a_1, a_2) = \textbf{Y}'_{T \land D}(n,a_1, a_2),$$ where $\textbf{Y}'$ stands for the collection of all derivatives of words in $\textbf{Y}$. As we show next, the first VT-type constraint in the preceding code $\textbf{Y}$ may be used to ``approximately'' correct the deletion and the adjacent transposition. Given that the approximate correction may be erroneous, the second VT-type constraint is used to perform exact correction.

We have the following lemma.

\begin{lemma} \label{lemma:equations}
For all $a_1, a_2 \in \mathbb{Z}_{n+2L+1}$, the code $\cC_{T \land D}(n,a_1, a_2)$ is a 1-TD code.
\end{lemma}
\begin{IEEEproof} We use the same approach as the one outlined in the proof of Claim~\ref{cl:eq}. 

Since $\sum_{i=1}^{n-1} (2i+1) x_i \equiv a_1 \bmod (n + 2L + 1)$ and $x_n=0$, we have that
\begin{align}\label{lem:1td1}
\sum_{i=1}^{n} i x'_i \equiv a_1 \bmod n+2L + 1.  
\end{align}
Furthermore, since $x_n = 0$,
\begin{align}\label{lem:1td2}
\sum_{i=1}^n x'_i \equiv 0 \bmod 2.
\end{align}
From (\ref{lem:1td1}) and (\ref{lem:1td2}), it is clear that if $\bfx \in \textbf{Y}_{T \land D}(n,a_1, a_2)$, then $\bfx' \in \cC_{VT}(n,a_1,0,L).$ Similarly to what was done in Theorem~\ref{th:general}, it can be shown that if $L \geq 1$ and $\textbf{Y}_{T \land D}(n,a_1, a_2)$ has Hamming distance at least $5$, then $\cC_{T \land D}(n,a_1, a_2)$ is a $1$-TD code. By design, $L \geq 1$ and so we turn our attention to showing that $\textbf{Y}_{T \land D}(n,a_1, a_2)$ has Hamming distance at least $5$.

We claim that the vectors in $\textbf{Y}_{T \land D}(n,a_1, a_2)$ represent a coset of a Berlekamp code \cite[Chapter 10.6]{Roth} with Lee distance $5$, which implies the desired result. 
To prove the claim, note that the binary code $\textbf{Y}_{T \land D}(n,0, 0)$ has a parity-check matrix of the form
\begin{align*}
H=\begin{bmatrix}
    3       & 5 & 7         & \dots & 2n-1 \\
    3^3   & 5^3 & 7^3 & \dots & (2n-1)^3
\end{bmatrix}.
\end{align*}
According to \cite[Chapter 10.6]{Roth}, in order for $\textbf{Y}_{T \land D}(n,0, 0)$ to have minimum Lee distance $5$, the following statement has to be true: For any two columns of $H$, say $h_{i}$ and $h_{j}$, it has to hold that
$$ h_{i_1} + h_{i_2} \neq \begin{bmatrix} 0 \\ c \end{bmatrix}, $$
for any possible choice of $c \in \mathbb{F}_{n+2L+1}$. Clearly, this condition is true since $n + 2L + 1$ is an odd prime and the sum of two odd numbers cannot equal another odd number. Thus, $\textbf{Y}_{T \land D}(n,0, 0)$ has minimum Lee distance at least $5$ and so $\textbf{Y}_{T \land D}(n,a_1, a_2)$ has minimum Lee distance at least $5$, as claimed.
\end{IEEEproof}
The above construction improves upon the general construction described by the result (\ref{eq:generaltrans}) in terms of $\log\,n$ bits of redundancy.

\begin{remark} It has been a long standing open problem to find extensions for the single-deletion VT code construction which would have \emph{order optimal} redundancy and impose syndrome constraints of the form $\sum_{i} f_k(i)\,x_i \equiv a\, \mod (n+1)$,
for some judiciously chosen functions $f_k(i)$. Attempts based on using this approach have failed so far~\cite{brakensiek2015efficient}. On the other hand, the result of Lemma~\ref{lemma:equations} shows that syndrome constraints of the form described above can accommodate combinations of one deletion and other forms of errors, such as adjacent transpositions.
\end{remark}

\begin{corollary} 
There exists a $1$-TD code which redundancy at most $2\log\,n+c$ bits, for some absolute constant $c$.
\end{corollary}

In the next section, we turn our attention to the problem of constructing codes capable of correcting transposition and deletion errors in the form of blocks of bits. First, we analyze the problem of constructing codes capable of correcting a single block of adjacent deletions. Then, we focus on constructing codes capable of correcting a single transposition of adjacent blocks in addition to handling one block deletion.

\section{Codes For Correcting a Block of Deletions}\label{sec:block}

We describe next a new family of codes capable of correcting one block of at most $b$ consecutive deletions; the codes require $\log b \, \log n + \mathcal{O}(b^2 \, \log\, b \, \log \log n)$ bits of redundancy, and hence improve upon the state-of-the art scheme which requires at least $(b-1) \log n$ bits of redundancy~\cite{schoeny2016codes}. The proposed block-deletion codes will subsequently be used in Section~\ref{sec:blockTransDel} to construct codes capable of correcting both a block of deletions (which we alternatively refer to a burst of deletions) and an adjacent transposition of two blocks of consecutive symbols. 

To explain the intuition behind our approach, we start with a short overview of existing code constructions for correcting a block of consecutive deletions, where the length of the block is fixed. It will be helpful to think of codewords of length $n=c\,b$, $c \geq 1$, as two dimensional arrays formed by writing the bits in the codeword column-wise, i.e., by placing the bits $(x_1,x_2, \ldots, x_b)$ in an orderly fashion within the first column of the array, the bits $(x_{b+1}, x_{b+2}, \ldots, x_{2b})$ within the second column and so on. As an example, for $c=n/b$, the codeword $\bfx = (x_1, \ldots, x_{n} )$ would read as follows:
\begin{align}\label{eq:cwmatrix}
\begin{bmatrix}
    x_{1}       & x_{b+1} & x_{2b+1} & \dots & x_{c(b-1) + 1} \\
    x_{2}       & x_{b+2} & x_{2b+2} & \dots & x_{c(b-1) + 2} \\
    \hdotsfor{5} \\
    x_{b}       & x_{2b} & x_{3b} & \dots & x_{n}
\end{bmatrix}.
\end{align} 
For simplicity, throughout the remainder of this section, we use the term ``interleaved sequence'' to refer to a row in the array. Note that in this setting, a block of $b$ consecutive deletions in a codeword $\bfx$ leads to one deletion within each interleaved sequence, and that the locations of deletions in the interleaved sequences are correlated. As an example, the block may cause the same deletion location in the first interleaved sequence, but affect the symbols in the other interleaved sequences differently (The deleted symbols are underlined):
\begin{align}\label{eq:cwmatrix}
\begin{bmatrix}
    x_{1}       & x_{b+1} & \underline{x}_{2b+1} & \dots & x_{c(b-1) + 1} \\
    x_{2}       & x_{b+2} & \underline{x}_{2b+2} & \dots & x_{c(b-1) + 2} \\
    \hdotsfor{5} \\
    x_{b}       & x_{2b} & \underline{x}_{3b} & \dots & x_{n}
\end{bmatrix},
\end{align} 
or
\begin{align}\label{eq:cwmatrix}
\begin{bmatrix}
    x_{1}       & x_{b+1} & \underline{x}_{2b+1} & \dots & x_{c(b-1) + 1} \\
    x_{2}       & \underline{x}_{b+2} & x_{2b+2} & \dots & x_{c(b-1) + 2} \\
    \hdotsfor{5} \\
    x_{b}       & \underline{x}_{2b} & x_{3b} & \dots & x_{n}
\end{bmatrix},
\end{align} 
or
\begin{align}\label{eq:cwmatrix}
\begin{bmatrix}
    x_{1}       & x_{b+1} & \underline{x}_{2b+1} & \dots & x_{c(b-1) + 1} \\
    x_{2}       & x_{b+2} & \underline{x}_{2b+2} & \dots & x_{c(b-1) + 2} \\
    \hdotsfor{5} \\
    x_{b}       & \underline{x}_{2b} & x_{3b} & \dots & x_{n}
\end{bmatrix}.
\end{align} 
As a result, by finding the location of the deletion in the first interleaved sequence does not automatically allow one to determine the ``shift'' of the block with respect to that location. Furthermore, deletion correcting codes such as VT codes only identify the \emph{run} of symbols in which the deletion occurs and not its exact position, as the goal is to reconstruct the correct codeword and not precisely determine the location of the error. As a result, further uncertainty exists about the locations of the deletions in the second, third etc. interleaved sequence of the codeword.

To mitigate these problems, the authors of~\cite{cheng2014codes} proposed a construction of codes capable of correcting a block of consecutive deletions of length \emph{exactly} $b$ based on imposing simple constraints on the interleaved sequences of a codeword. A construction with redundancy of approximately $b \log n$ bits requires \emph{all} the interleaved sequences of (\ref{eq:cwmatrix}) to belong to a VT code. The main drawback of this construction is that each interleaved sequence is treated independently of the others and that consequently, the redundancy of the codes is too high. To address this problem, one should use the position of the deletion in the first interleaved sequence to approximately determine the location of the deletion in the second row and similarly for all other subsequent rows. In~\cite{cheng2014codes}, the authors also proposed a code which has an alternating sequence (i.e., a sequence of the form $0,1,0,1,0,1,\ldots$) as its first interleaved sequence and all the remaining interleaved sequences satisfying a constraint that requires $\log 3$ bits of redundancy. The proposed code may be easily decoded by first determining the location of the deletion in the first {row} through a reference to the alternating sequence structure. Then, this location is used by the remaining rows to correct the remaining $b-1$ deletions. This approach requires at least $n/b$ bits of redundancy, due to the fact that one has to fix the first row of the codeword array. Thus, the redundancy of this approach is actually higher than that of the code that uses individual VT code constraints for each interleaved sequence. 

The alternating sequence approach was improved and generalized in~\cite{schoeny2016codes}, where the authors constructed block deletion-correcting codes with a significantly more relaxed constraint placed on the first interleaved sequence. Their idea was to combine constrained coding with a variant of VT codes which we explain in details in what follows. The relaxed constraints allow one to \emph{approximately} determine the locations of the remaining deletions in $\bfx$ after decoding the first interleaved sequence of the array. The constrained and VT-type constraints imposed on the higher index rows nevertheless allow for unique recovery of the codeword $\bfx$ by using VT codes confined to the ``suspect range'' predicted to harbor the deletions. The codes constructed in~\cite{schoeny2016codes} require approximately $\log n$ bits of redundancy for the constraint in the first row of the array, and $\log \log n$ bits of redundancy for each of the remaining rows. This results in a total redundancy of roughly $\log n + (b-1) \log \log n$ bits for correcting a block of consecutive deletions of length exactly $b$ (compared to the redundancy of~\cite{cheng2014codes} which equals $b \log n$ bits). 

To allow for correcting any single block of length at most $b$, the codes from~\cite{schoeny2016codes} have to be changed so as to include nested redundant bits that capture multiple coding constraints and may allow for correcting a range of block lengths. Which of the constraints to use is apparent upon observing the length of the received word: To correct one block of any possible length at most $b$, the decoder for the underlying code locates the position of the block of consecutive deletions differently for each possible block length. For instance, if $\bfx$ experiences a block error of length $b_1 \leq b$, then the code uses one VT-type constraint, say $K_{VT,b_1}$. 

However, if $\bfx$ experiences an error burst of length $b_2$ with $b_2 <b_1$, then the code effectively uses a different VT-type constraint, say $K_{VT,b_2}$. Note that since each of the constraints $K_{VT,b_i}, \, 2 \leq i \leq b,$ is de facto a VT-type constraint, one requires roughly $(b-1) \log n + b^2 \log \log n$ bits of redundancy, compared to the $b^2 \log\, n$ redundancy which would have been required by the scheme in~\cite{cheng2014codes}.

Our approach in this work for a further improvement is to reuse the same VT-type constraint for multiple possible block lengths, in which case the redundancy will amount to roughly $\log b \, \log n + \log b \, b^2 \log \log n$ bits. To describe this method, we start with a construction that allows for correcting one odd-length block of consecutive deletions of length at most $b$, and then proceed to extend the result for even-length blocks.

\subsection{Odd Length Blocks}\label{subsec:oddB}

Our code construction is centered around three main ideas:
\begin{enumerate}
\item The use of VT codes (\ref{eq:VTconstraint}).
\item The use of running sum constraint (\ref{eq:codingc}).
\item The use of a sequence of \emph{Shifted VT codes}~\cite{schoeny2016codes}, defined in (\ref{eq:SVT1}) i.e., codes that enforce multiple modular VT-type constraints with parameter values smaller than $n+1$.  
\end{enumerate}
As discussed in more details in what follows, our choice of the {Shifted} VT codes requires approximately $b^2 \log \log n$ bits  of redundancy and the constrained coding constraint requires a single bit of redundancy, the proposed construction introduces  roughly $\log n + b^2 \log \log n $ bits of redundancy. 

The decoder operates as follows. Suppose that $\bfy$ is the result of deleting $t$ consecutive bits from $\bfx$, with $t \leq b$ and $t$ odd. Then,
\begin{enumerate}
\item The decoder computes a number of parities and decides on the appropriate Shifted VT code (\ref{eq:SVT1}) to use in determining the Hamming weight of the bits deleted from $\bfx$. 
\item Using both the VT-type constraint (\ref{eq:VTconstraint}) and the constraint (\ref{eq:codingc}), the decoder determines an approximate location for the block deletion in $\bfx$ that resulted in $\bfy$. 
\item Given the approximate location of the block of deletions, the decoder uses a series of Shifted VT codes (\ref{eq:SVT1}) to determine the exact locations and values of the bits deleted from $\bfx$ that lead to $\bfy$.
\end{enumerate}

\textbf{Part 1. Determining the weight of the deleted substring.} We start with some relevant terminology and notation. For a word $\bfx \in \mathbb{F}_2^n$, let $\cB_{D,\leq b}(\bfx)$ denote the set of all words that may be obtained from  $\bfx$ by deleting at most $b$ consecutive bits. For example, for $\bfx=(0,1,1,0,0,1) \in \mathbb{F}_2^6$, we have 
\begin{align*}
\cB_{D,\leq 2}(\bfx) = \Big \{ &(0,1,1,0,0,1), (1,1,0,0,1), (0,1,0,0,1), \\
 &(0,1,1,0,1), (0,1,1,0,0), (1,0,0,1), (0,0,0,1), \\
&(0,1,0,1),(0,1,1,1), (0,1,1,0) \Big \}. 
\end{align*}
Similarly, let $\cB_{D,b}(\bfx)$ denote the set of words that may be obtained from $\bfx$ by deleting \textit{exactly} $b$ consecutive bits. 

Furthermore, given a vector $\bfd \in \mathbb{F}_2^{b}$, define the code $\cC_{par}(n,b, \bfd)$ as follows\footnote{We use the subscript $par$ to refer to the function of the code, which is to recover the weight of the deleted block (substring) by using a parity check.}:

\begin{align*}
\cC_{par}(n,b,\bfd) = \{ \bfx \in \mathbb{F}_2^n : \forall j \in [b], \sum_{i=0}^{\lfloor \frac{n-j}{b} \rfloor} x_{j+b i} \equiv d_j \bmod 2 \}.
\end{align*}
It is straightforward to see that the code imposes a single parity-check constraint on the interleaved sequences of $\bfx$, which suffices to determine the weight of the deleted block. In addition, we observe that we used a set of parameters $d_i$ for the weight constraints, rather than the classical even parity constraints for reasons that will become apparent in the subsequent exposition. In a nutshell, the resulting codes of the section will be nonlinear and averaging arguments for the size of codes require the use of a range of parameter values.

\begin{example}\label{ex:weightd} Suppose that $\bfx = (0,1,1,0,0,1, 0,1,0,1,0,1) \in \cC_{par}(12,2,(1,1))$ was transmitted and $\bfy = (0,1,1,0,0,1,0,1,0,1)$ $\in \cB_{D,2}(\bfx)$ was received instead. Since $\bfx \in \cC_{par}(12,2,(1,1))$, it is straightforward to determine that the bits $0,1$ were deleted from $\bfx$ to obtain $\bfy$. Notice, however, that we cannot infer the order in which the deleted bits $\{0,1\}$ appeared in $\bfx$ from the constraints of the code $\cC_{par}(12,2,(1,1))$, nor their exact location.
\end{example}

\textbf{Part 2. Imposing the generalized VT conditions.}

Given $\bfy \in \mathbb{F}_2^{n-b}$, $\bfv \in \mathbb{F}_2^b$ and $k_I \in [n-b+1]$, let $I(\bfy, \bfv, k_I) \in \mathbb{F}_2^n$ be the vector obtained by inserting $\bfv$ into $\bfy$ at position $k_I$. For instance, if $\bfy= (0,1,1,0,0,1) \in \mathbb{F}_2^6$, $k_I=1$ and $\bfv=(0,1)$, then $I(\bfy, \bfv, 1) = (0,1,0,1,1,0,0,1) \in \mathbb{F}_2^8$. Similarly, let $D(\bfy,b,k_D)$ be the result of deleting $b$ consecutive bits from $\bfy$ starting at position $k_D$. Thus, $D(\bfy, 2,2) = (0,0,0,1) \in \mathbb{F}_2^4$. As before, let $wt(\bfx)$ stand for the Hamming weight of a vector $\bfx$.

\begin{claim}\label{cl:vtBurstD} Let $\bfv_1,\bfv_2 \in \mathbb{F}_2^b$, and suppose that $\bfy \in \mathbb{F}_2^{n-b}$ and that $w_1=wt(\bfv_1)=wt(\bfv_2)$, $w_2 = wt(y_{i_1}, \ldots, y_{i_2-1})$, where $i_1<i_2$. Let ${\bfx} = I(\bfy, \bfv_1, i_1) \in \mathbb{F}_2^n$ and $\bfu = I(\bfy, \bfv_2, i_2) \in \mathbb{F}_2^n$. Then,
\begin{align} \label{eq:difference}
\sum_{i=1}^n i  u_i - \sum_{i=1}^n i x_i = (i_2-i_1) w_1 - b w_2 + \delta,
\end{align}
where $|\delta| < b^2$.
\end{claim}
\begin{remark} Note that the term $\delta$ arises due to the fact that the sequences $\bfv_1$ and $\bfv_2$ have the same weight but potentially different locations of their nonzero symbols.
\end{remark}

Let $\cC_{VT,b}(n,a,b)$ be a VT code of the form 
\begin{align}\label{eq:VTconstraint}
\cC_{VT,b}(n,a,b) = \{ \bfx \in \mathbb{F}_2^n : \sum_{i=1}^n i  x_i \equiv a \bmod (b n+b^2) \}.
\end{align}
Clearly, two vectors $\bfx, \bfu \in \mathbb{F}_2^n$ of the form as defined in Claim~\ref{cl:vtBurstD} cannot both lie in the same code capable of correcting a block of deletions of length at most $b$. To see this, note that $\bfy=D(\bfx, b,i_1) = D(\bfu, b, i_2) \in \cB_{D,b}(\bfx) \cap \cB_{D,b}(\bfu),$ a contradiction.

If we assume that $\bfx, \bfu \in \cC_{VT,b}(n,a, b)$ are typical sequences generated by an i.i.d uniform source, 
then, with high probability, $w_2$ will be close in value to $\frac{i_2-i_1}{2}$. 
In order for $\bfx$ and $\bfu$ to belong to different codes (i.e., codes with different VT syndromes) capable of correcting block deletions of length at most $b$ for an overwhelming large portion of the constituent vectors $\bfv$, based on Claim~\ref{cl:vtBurstD} and the definition of $\cC_{VT,b}(n,a, b)$, we need to ensure that the right-hand side of~(\ref{eq:difference}) is not zero, i.e., that 
\begin{align}\label{eq:compareBO}
 (i_2-i_1) w_1 + \delta \neq  b \, \frac{i_2-i_1}{2}.
\end{align}
Note that if $b$ is odd, then $w_1$ cannot be equal to $b/2$.

Next, we construct a codebook that ensures that (\ref{eq:compareBO}) is satisfied for any choice of distinct (code)words. The idea is to construct a set of codewords $\bfx$ with the following property: Every block (substring) in $\bfx$ of length $i_2-i_1=B,$ where $B \geq b^4 \log n,$ is required to have Hamming weight approximately equal to $\frac{B}{2}$. If every block of $B \geq b^4 \log n$ bits in $\bfx$ has weight close to $B/2$ and if $b$ is odd, then one can show (see Lemma~\ref{lem:burstVT}) that for any $i_1,i_2$ such that $i_2-i_1 \geq B$, $\bfx, \bfu$ do not simultaneously belong to the same VT code.

Therefore, when a block of deletions occurs, Lemma~\ref{lem:burstVT} shows that it will be possible to determine approximately (to within $B$ bits) the location of the block of deletions by attempting to insert a block of bits into different positions and check whether they lead to a vector that satisfies a VT-type constraint. According to Lemma~\ref{lem:burstVT}, if the VT-type constraint is satisfied, we know the location of the block of deletions to within $B$ positions and we can determine exactly the value and location of the deleted bits using the Shifted VT codes in (\ref{eq:SVT1}).

For notational convenience, we henceforth assume that $n$ is a power of two so that $\log n$ is a positive integer.

Let $Bal(n,b)$ denote the following ``balanced'' set of sequences: 
\begin{align}
Bal(n,b) = \Big \{& \bfx \in \mathbb{F}_2^n : \forall B \in [n], B \geq b^4 \log n,  \forall j \in [n-B+1], \nonumber\\
 & \frac{B}{2} - \frac{B}{3b} < \sum_{i=j}^{j+B-1} x_i < \frac{B}{2} + \frac{B}{3b} \Big \}. \label{eq:codingc}
\end{align}

We have the following claim, the proof of which may be found in Appendix~\ref{App:clCSB}.

\begin{claim}\label{cl:CSB} For a positive integer $n \geq 10$ and $b \geq 5$, 
$$\log |Bal(n,b)| \geq n + \log  \left( 1-2 \, n^{2-\frac{2}{9}b^2\log e} \right).$$ 
\end{claim}

Note that as a consequence of Claim~\ref{cl:CSB}, we have 
\begin{align}\label{eq:Sndredudancy}
\log |Bal(n,b)| \geq n-1,
\end{align}
and the coding constraint incurs not more than one bit of redundancy.

Let $\bfD = (\bfd_1, \bfd_2, \ldots, \bfd_b),$ where for $i \in [b]$, $\bfd_i=(d_{1,i}, \ldots, d_{i,i}) \in \mathbb{F}_2^i$. Define
\begin{align*}
\cC^{odd}_b(n,a, \bfD) = \{ \bfx \in \mathbb{F}_2^n : &\sum_{i=1}^n i  x_i \equiv a \bmod bn + b^2, \\
&\bfx \in Bal(n,b), \forall i \in [b] \\
&\bfx \in \cC_{par}(n, i, \bfd_i) \}.
\end{align*}
At a high level, the parity constraint $\cC_{par}$ is used to determine the weight of the deleted block of symbols, and, similar to the previous discussion, the combination of the VT-type constraint along with the balancing constraint ($Bal(n,b)$) allows one to approximately determine the location of the deletions. In particular, the balancing constraint prohibits VT decoding errors, while the parity constraint is imposed on all interleaved sequences, as dictated by the vectors $\bfd_i$, $i \in [b]$.

We show next that if $\bfy$ is the result of an odd-length block of deletions occurring in $\bfx \in \cC^{odd}_{b}(n,a,\bfD)$ starting at position $k_D$, then there exists a decoder for $\cC^{odd}_{b}(n,a,\bfD)$ that is capable of producing an estimate, say $\hat{k}_D$, for the starting position of the block of deletions, with $|k_D - \hat{k}_D|< b^4 \log n$. We then proceed to explain how to recover the exact value and location of the deleted bits using the Shifted VT codes of (\ref{eq:SVT1}). 

First, we show that if {$\bfy \in\cB_{D,t}(\bfx)$} and $\bfx \in \cC_{b}^{odd}(n,a, {\bfD}),$ with $t \leq b$ an odd integer, one can obtain a good estimate for the location of the block of deletions given that we know the Hamming weight of the bits that were deleted (we can obtain this from a decoder for the subcode $\cC_{par}(n,t,\bfd_t)$). The following result, similar to Claim~\ref{cl:vtBurstD}, describes the relevant properties of the decoder. 

\begin{lemma}\label{lem:burstVT}  Let $\bfv_1,\bfv_2 \in \mathbb{F}_2^{t}$, $wt(\bfv_1) = wt(\bfv_2)$, $\bfy \in \mathbb{F}_2^{n-t}$, and suppose that $t$ is an odd number where $t \leq b$ and $b \geq 5$.  For $i_1<i_2$ and $i_2-i_1 \geq b^4 \log n$, let ${\bfx} = I(\bfy, \bfv_1, i_1) \in \cC^{odd}_{b}(n,a, \bfD)$ and $\bfz = I(\bfy, \bfv_2, i_2) \in \mathbb{F}_2^n$. Then,
\begin{align*}
\sum_{i=1}^n i \, z_i - \sum_{i=1}^n i \, x_i \not \equiv 0 \bmod b\, n + b^2,
\end{align*}
and hence $\bfz \not \in \cC^{odd}_b(n,a, \bfD)$.
\end{lemma}
\begin{IEEEproof} As before, let $w_1=wt(\bfv_1) = wt(\bfv_2)$ and $w_2 = wt(y_{i_1}, \ldots, y_{i_2-1})$. 
Now according to Claim~\ref{cl:vtBurstD}, 
$$\sum_{i=1}^n i  z_i - \sum_{i=1}^n i x_i =B  w_1 - t \,  w_2 + \delta,$$ and therefore our goal is to show that
\begin{align}\label{eq:lem15mainid}
B  w_1 + \delta \not \equiv t \,  w_2 \bmod bn + b^2,
\end{align}
whenever $B =i_2 - i_1 \geq b^4 \log n$, which will establish the statement of the lemma.

Since $\bfx \in \cC_{b}^{odd}(n,a, \bfD)$, one has $\bfx \in Bal(n,b)$ and hence it follows from (\ref{eq:codingc}) that 
$$\frac{B}{2} - \frac{B}{3b}< \sum_{i=i_1}^{i_2-1} y_i < \frac{B}{2} + \frac{B}{3b}.$$ 
Thus, given that $t \leq b$, 
\begin{align*}
\frac{B \, t}{2} - \frac{B}{3} < t \, w_2 < \frac{B \, t}{2} + \frac{B}{3}.
\end{align*}
Notice that since $t$ is odd, and since $w_1 = (t+k)/2,$ where $-b \leq k \leq b$, $k$ is odd. Thus, we have
\begin{align*}
B w_1 + \delta = \frac{B\,t}{2} + k \, \frac{B}{2} + \delta,
\end{align*}
where $k \neq 0$. We will prove the result for the case when $k$ is positive. 
The result may be proved similarly for negative $k$.
 
For $k \geq 1$, we have 
$$B w_1 + \delta \geq \frac{B \, t}{2} + \frac{B}{2} - b^2.$$ 
Since $B \geq b^4 \log n$ and $b \geq 5$, it follows that  
$$B w_1 + \delta \geq \frac{B\, t}{2} + \frac{B}{2} - b^2 > \frac{B \, t}{2} + \frac{B}{3} > t\,  w_2, $$
and hence (\ref{eq:lem15mainid}) holds.
\end{IEEEproof} 

The desired code $\cC^{odd}_{b}(n,a,\bfC, \bfD) \subseteq \mathbb{F}_2^n$, capable of correcting any single block of odd length $t \leq b$, is a subcode of the code $\cC^{odd}_{b}(n,a,\bfD) \subseteq \mathbb{F}_2^n$. From Lemma~\ref{lem:burstVT}, we know that the code $\cC^{odd}_{b}(n,a,\bfD)$ can approximately determine the location of the block of deletions, assuming the block is odd. In what follows, we describe Shifted VT codes which will be used in Part 4 to exactly pinpoint the locations and values of the deleted symbols.

\textbf{Part 3. Incorporating Shifted VT codes.} We now briefly turn our attention to Shifted VT codes introduced in~\cite{schoeny2016codes}. For completeness, we state the results necessary for our subsequent derivations and provide an example of the decoding process. For further details, see Appendix~\ref{App:SVTProp}.

The Shifted VT code with positive integer parameters $c,d$ and $n,M$, denoted $SVT_{c,d}(n, M)$, is defined as follows:

\begin{align}
SVT_{c,d}(n,M) = \{ \bfx \in \mathbb{F}_2^n :& \sum_{i=1}^n i \, x_i \equiv c \bmod \, M,  \nonumber \\
&\sum_{i=1}^n x_i \equiv d \bmod 2 \}. \label{eq:SVT1}
\end{align}

A Shifted VT code is capable of determining the exact location and value of a bit deleted from a codeword provided some sufficiently accurate estimate for the location of the deletion is known. To see why this is true, observe that the modulus of the sum in the definition equals $M$, which is assumed to be significantly smaller than $n+1$, the modulus used in classical VT codes. The code basically imposes a VT-type constraint, which can correct a single deletion error, but on a substring of the sequence. This code property is described more precisely in the next lemma.

\begin{lemma}\label{lem:SVTBasics} Suppose that $\bfy \in D(\bfx, 1, k_D)$, where $\bfx \in SVT_{c,d}(n,M)$ and where $M \geq 2P-1$, $d_b \in \mathbb{F}_2$. Given a $\hat{k}_D$ such that $|k_D - \hat{k}_D| < P$, there exists at most one possible value for $k_D'$ and for $\bfd_b$ that jointly satisfy $I(\bfy, d_b, k_D') \in SVT_{c,d}(n,M)$. In this setting, we have $I(\bfy, d_b, k_d') = \bfx$.  \end{lemma}

\begin{example} Suppose that $\bfx = (0,1,1,0,0,1) \in SVT_{2,1}(6,2\, P - 1)$, $\bfy = D(\bfx, 1, 3) = (0,1,0,0,1)$,  $\hat{k}_d = 4$, and $P=2$. Note that in the example, $k_d = 3$ so that $|\hat{k}_d - k_d| < P$, as required by the setup of Lemma~\ref{lem:SVTBasics}.

First, we can determine that the value of the deleted bit is $1,$ given that $\bfy$ and $\sum_{i=1}^6 x_i \equiv 1 \bmod 2$. We may also conclude that $k_d \in \{ 3,4,5\},$ since the decoder for $SVT_{2,1}(6,2\, P - 1)$ provided the estimate $\hat{k}_d=4,$ and we already had the prior knowledge that $|\hat{k}_d - k_d| < P$. To proceed, we need to examine each of the following three potential deletion locations:
\begin{align*}
\hat{\bfx}_1 =& (0, 1, 1, 0, 0, 1), \text{for $k_d=3$,}\\
\hat{\bfx}_2 =& (0, 1, 0, 1, 0, 1), \text{for $k_d=4$,} \\
\hat{\bfx}_3 =& (0, 1, 0, 0, 1, 1), \text{for $k_d=5$.}
\end{align*}
Observe that $\hat{\bfx}_1 \in SVT_{2,1}(6,3)$, $\hat{\bfx}_2 \in SVT_{0,1}(6,3)$, $\hat{\bfx}_3 \in SVT_{1,1}(6,3)$. Thus, the decoder for $SVT_{2,1}(6,3)$ can conclude that $\bfx = \hat{\bfx}_1,$ since this is the only one of the three vectors that belongs to the code $SVT_{2,1}(6,3)$.
\end{example}

\textbf{Part 4. Combining the different code construction components.} Next, we describe a family of codes $\cC^{odd}_{b}(n,a, \bfC, \bfD)$, capable of correcting any odd-length block of consecutive deletions of length not exceeding $b$. 

Let $\bfD = (\bfd_1, \bfd_2, \ldots, \bfd_b),$ where for $i \in [b]$, $\bfd_i=(d_{1,i}, \ldots, d_{i,i}) \in \mathbb{F}_2^i$. Furthermore, let $\bfC = ( \bfc_1, \bfc_2, \ldots, \bfc_b)$ where for $i \in [b]$, $\bfc_i=(c_{1,i}, \ldots, c_{i,i}) \in \mathbb{Z}^i_{2 b^5 \log n}$. For a vector $\bfx \in \mathbb{F}_2^n$, let $\bfx^{(f,b)}$ be the $f$-th interleaved sequence of $\bfx$. For instance if $\bfx = (0,1,1,0,0,1)$, then $\bfx^{(1,2)}=(0,1,0)$. Similarly, $\bfx^{(2,2)} = (1,0,1)$. 

We {define} a code $\cC^{odd}_{b}(n,a,\bfC, \bfD) \subseteq \mathbb{F}_2^n$ as follows\footnote{The superscript $odd$ is used to indicate the fact that the length of the block of deletions is odd.}: 
\begin{align}\label{eq:oddburst}
\cC^{odd}_{b}&(n,a, \bfC, \bfD) = \Big \{ \bfx \in \mathbb{F}_2^n : \sum_{i=1}^n i x_i \equiv a \bmod bn + b^2,  \nonumber \\
& \bfx \in Bal(n,b), \text{ and } \forall i_2 \in [b], \forall i_1 \leq i_2, \nonumber \\  
&\bfx^{(i_1,i_2)} \in SVT_{{c_{i_1,i_2}},{d_{i_1,i_2}}}( \lfloor \frac{n-i_1}{i_2} \rfloor + 1, 2 (b^5 \log n + b)  \Big \},
\end{align}
where the $d_{i,j}$s are elements of the vectors $\bf{d}_i$ defined above.

Note that $\cC^{odd}_{b}(n,a, \bfC, \bfD) \subseteq \cC^{odd}_b(n,a, \bfD),$ and in particular, if $\bfx \in SVT_{{c_{i_1,i_2}},{d_{i_1,i_2}}}(n, 2 (b^5 \log n + b))$ for $i_1 \leq i_2$, then $\bfx \in \cC_{par}(n, i_2, {\bfd_{i_2}})$. As before, the constraints imposed by the code $\cC^{odd}_{b}(n,a, \bfD)$ (the VT-type constraint, the balancing constraint, and the parity constraint) enable one to approximately determine the location of the deletions. Given this information, the SVT constraints is used to correct the block of deletions. The indices $i_1,i_2$ describe the locations/lengths of the substrings of $\bfx$ on which the Shifted VT constraint is imposed, while the parameters $\bfC, \bfD$ specify the VT-modulus and parity constraints of the Shifted VT code, respectively.

\begin{theorem}\label{th:oddburst} Suppose that $\bfx \in \cC_b^{odd}(n,a,\bfC, \bfD)$ and that $\bfy \in \cB_{D,t}(\bfx),$ where 
$t$ is an odd integer such that $t \leq b$, $b \geq 5$. Then, there exists a decoder for $\cC_b^{odd}(n,a,\bfC, \bfD)$ that can recover $\bfx$ from $\bfy$. \end{theorem}
\begin{IEEEproof}
Assume that $\bfy = D(\bfx, t, k_D)$ and let $\bfv = (x_{k_D}, x_{k_{D}+1}, \ldots, x_{k_D+t-1})$. First, we use the fact that $\bfx \in \cC_{par}(n, t, {\bfd_t}),$ which follows from the constraint that $\bfx \in  SVT_{{c_{i_1,i_2}},{d_{i_1,i_2}}}(\lfloor \frac{n-i_1}{i_2} \rfloor +1, 2 (b^5 \log n + b),$ 
to determine the precise values of the deleted bits. For this purpose, let $w$ denote the number of deleted nonzero symbols. Clearly, $wt(\bfv) = w$.

Next, we determine $\hat{\bfv} \in \mathbb{F}_2^t$ and a $\hat{k}_D \in [n-t+1]$ such that $wt(\hat{\bfv}) = w$ and $I(\bfy, \hat{\bfv}, \hat{k}_D) \in \cC_b^{odd}(n,a,\bfD)$. Since $I(\bfy, \bfv, k_D) \in \cC_b^{odd}(n,a, \bfC, \bfD)$ and $wt(\bfv) = wt(\hat{\bfv})$, it follows from  Lemma~\ref{lem:burstVT} that if $I(\bfy, \hat{\bfv}, \hat{k}_D) \in \cC_b^{odd}(n,a,\bfD)$, then $|k_D - \hat{k}_D| < b^4 \log n$. Finally, we use the constraint $\bfx \in SVT_{{c_{i_1,i_2}},{d_{i_1,i_2}}}(\lfloor \frac{n-i_1}{i_2} \rfloor +1, 2 (b^5 \log n + b))$ once again to recover the exact locations and values of the deleted bits.
\end{IEEEproof}

\begin{example} Suppose that $\bfx=(0,1,1,0,0,0,1,1,0,0,1,1,0)$, so that $\bfx \in \cC_{par}(13,3,(1,1,0))$ and $\sum_{i=1}^{13} i \, x_i \equiv 43 \bmod 48$. If $\bfy=(0,1,1,1,1,0,0,1,1,0) \in \cB_{D,3}(\bfx)$, then there exists only one vector $\hat{\bfx}=(\hat{x}_1, \ldots, \hat{x}_n)$ such that $\hat{\bfx} \in \cC_{par}(13,3,(1,1,0))$ and $\sum_{i=1}^{13} i \, x_i \equiv 43 \bmod 48$, namely $\hat{\bfx} = \bfx$.
\end{example}

\begin{corollary}\label{cor:oddb} Let $\bfx, \bfu \in \cC_b^{odd}(n,a, \bfC, \bfD) \subseteq \mathbb{F}_2^n,$ where $\bfx \neq \bfu$ and let $t$ be an odd integer such that $t \leq b$ and $b \geq 5$. Then, $\cB_{D,t}(\bfx) \cap \cB_{D,t}(\bfu) = \emptyset$, and for any $n \geq 10$, $\cC_b^{odd}(n,a, \bfC, \bfD)$ has at most
\begin{align*}
&\log(bn + b^2) + \frac{b(b+1)}{2}\Bigg( \log \Big( 2 (b^5 \log n + b) \Big) + 1 \Bigg) + 1
\end{align*}
bits of redundancy.
\end{corollary}
The result concerning the cardinality of the code follows from a straightforward averaging argument. The proof of the result can be found in Appendix~\ref{app:pfcoroddb}.

\subsection{The general case}\label{sec:general}

We now turn our attention to extending the previous construction so that it applies to blocks of arbitrary length -- odd or even -- not exceeding $b$.
The gist of the approach is to decompose a block of length $b$ into odd blocks, when viewed through the interleaved sequences of $\bfx$. The next example illustrates how this will be accomplished.

\begin{example}\label{ex:general} Suppose that $\bfx = (0,0,1,0,1,0,0,1,0) \in \mathbb{F}_2^9$ is transmitted and that the vector $\bfy = (0,1,0) \in \cB_{D, 6}(\bfx)$ is received instead. Notice that in this case, the sequence $\bfx^{(1,2)} = (0,1,1,0,0)$ experienced an odd block of consecutive deletions of length three, resulting in $\bfy^{(1,2)} = (0,0)$.  
\end{example}

Define the codebook $\cC_b(n,\bfa, \vec{\bfC},\vec{\bfD})$ according to
\begin{align}
\cC_b(n,\bfa, \vec{\bfC}&, \vec{\bfD}) = \Big \{ \bfx \in \mathbb{F}_2^n : \nonumber \\
&\forall j \in [\lceil \log b \rceil ], \bfx^{(1,2^{j-1})} \in \cC^{odd}_{\tilde{b}}(\lceil \frac{n}{2^{j-1}} \rceil , a_j, \bfC_j, \bfD_j) \nonumber \\
&\text{where } \tilde{b} = \max \{ \lceil \frac{b}{2^{j-1}} \rceil, 5 \}  \Big \}, \label{eq:generalBurst}
\end{align}
where $\bfa = (a_1, \ldots, a_{\lceil \log b \rceil})$, $\vec{\bfC} = (\bfC_1, \ldots, \bfC_{\lceil \log b \rceil})$, $\vec{\bfD} =( \bfD_1, \ldots, \bfD_{\lceil \log b \rceil})$, and where the codes $\cC^{odd}_{\tilde{b}}(\lceil \frac{n}{2^{j-1}} \rceil, a_j, \bfC_j, \bfD_j)$ are as defined in (\ref{eq:oddburst}). Based on Theorem~\ref{th:oddburst}, the fact that $\bfx^{(1,2^{j-1})} \in \cC^{odd}_{\tilde{b}}(\lceil \frac{n}{2^{j-1}} \rceil , a_j, \bfC_j, \bfD_j)$ implies that if $\bfx^{(1,2^{j-1})}$ experiences a burst of deletions of odd length, one can still correctly recover  $\bfx^{(1, \textcolor{black}{2^{j-1}})}$. As described next, $\bfx^{(1, \textcolor{black}{2^{j-1}})}$ is used to produce an estimate for the location of the burst of deletions in $\bfx$.

Let $\bfy= D(\bfx, t, k_D) \in \cB_{D, \leq b}(\bfx)$. Similarly to what was done in the context of odd blocks, we first produce an estimate $\hat{k}_D$ for ${k}_D$. 
Suppose that $t = 2^{j-1} \, (2l+1)$. We use the constraint $\bfx^{(1,2^{j-1})} \in  \cC^{odd}_{\tilde{b}}(\lceil \frac{n}{2^{j-1}} \rceil , a_j, \bfC_j, \bfD_j)$ to determine the sequence $\bfx^{(1,2^{j-1})}$, and use this information to compute $\hat{k}_D$. Subsequently, using $\hat{k}_D$ and the Shifted VT code constraints, we can determine the correct locations and values of the deleted bits.  

We say that $\bfv \in \mathbb{F}_2^B$ is a $b$-repeating pattern of length $B$ if $b | B$, and for any $1 \leq k \leq B/b -1$ we have $(v_1, \ldots, v_b) = (v_{bk+1}, \ldots, v_{bk+b})$. For instance $\bfv = (0,1,1,0,1,1)$ is a $3$-repeating pattern of length $6$. We find the following claim useful for the proof of the main result of this section.

\begin{claim}\label{cl:reppat} Suppose that $\bfv \in \mathbb{F}_2^B$ is a $b$-repeating pattern of length $B$, with $b$ odd, that appears as a substring in $\bfx \in Bal(n,b)$. Then, $B < b^4 \log n$. \end{claim}
\begin{IEEEproof} The result claims that self-repeating patterns in codewords $\bfx$ of the code under consideration have to be sufficiently short. To prove the claim, suppose that on the contrary, there exists a $b$-repeating pattern $\bfv \in \mathbb{F}_2^B$ in $\bfx \in Bal(n,b)$ such that $B \geq b^4 \log n$. Let $wt(v_1, \ldots, v_b) = (b+k)/2$ where, since $b$ is odd, $k$ is odd. In particular, $k \neq 0$. Then,
$$ \sum_{i=1}^B v_i = \frac{\frac{b+k}{2}}{b} \, B = \frac{B}{2} + \frac{Bk}{2b}, $$
and we arrive at a contradiction since in this case $\bfv$ cannot be a substring of $\bfx \in Bal(n,b)$.
\end{IEEEproof}
The $b$-repeating patterns serve the same role as runs in the single deletion case when applied to a block of consecutive deletions. Hence, a VT-type code can only determine in which $b$-repeating pattern the deletions occurred, but not the exact position of the block. This observation is illustrated by the following example.

\begin{example} Suppose that the vector $\bfx = (0,1,1,0,1,1,0,0,1)$ $\in \cC$ was transmitted and that the vector $\bfy = (0,1,1,0,0,1)$ was received. Note that given $\bfx$ and $\bfy,$ it is possible to determine that the substring $(0,1,1)$ was deleted from $\bfx$ to generate $\bfy$. However, observe that $\bfx = I(\bfy, (0,1,1), 1) = (\bfy, (0,1,1), 4),$ where both positions $1$ and $4$ are contained within a $b$-repeating pattern of length $6$.
\end{example}

We are now ready to state the main result of the section. 

\begin{theorem}\label{th:burstcode} Suppose that $\bfx \in \cC_b(n,\bfa, \vec{\bfC}, \vec{\bfD}),$ where $b \geq 5$ is odd, is transmitted, and that $\bfy \in \cB_{D, \leq b}(\bfx) = D(\bfx, t, k_D),$ where $1 \leq t \leq b$, is received instead. Then there exists a decoder for $\cC_b(n,\bfa, \vec{\bfC}, \vec{\bfD})$ capable of uniquely determining $\bfx$ from $\bfy$.
\end{theorem}
Note that the assumption that $b$ is odd is made for simplicity of analysis, and that all number of errors $t \leq b$ may be corrected
independent on their parity.
\begin{IEEEproof} Suppose that $\bfy$ has length $n-t,$ where $t$ is an odd integer. Then, the result immediately follows from the fact that $\bfx =\bfx^{(1,2^{0})} \in \cC^{odd}_{b}(n,a_1, \bfC_1, \bfD_1)$ and Theorem~\ref{th:oddburst}. 

Suppose next that $t = 2^{j-1} (2l+1)$ for some positive integer $j$ and $l \geq 0$.  Since 
$$\bfx \in \cC_{\tilde{b}}(n,\bfa, \vec{\bfC}, \vec{\bfD}),$$ 
we know that 
$$\bfx^{(1,2^{j-1})} \in \cC^{odd}_{\tilde{b}}(\lceil \frac{n}{2^{j-1}} \rceil , a_j, \bfC_j, \bfD_j),$$ 
where $j$, $\tilde{b}$ are as stated in the claim. Thus, it is possible to determine $\bfx^{(1,2^{j-1})}$ from $\bfy^{(1,2^{j-1})}$ since $\bfy^{(1,2^{j-1})} \in \cB_{D,2l+1}(\bfx^{(1,2^{j-1})})$. Note that from $\bfx^{(1,2^{j-1})}$ and $\bfy^{(1,2^{j-1})}$, 
we can determine the $(2l+1)$-repeating pattern in which the deletions occurred in $\bfx^{(1,2^{j-1})}$ so as to produce $\bfy^{(1,2^{j-1})}$.

Assume now that $\bfy^{(1,2^{j-1})} = D(\bfx^{(1,2^{j-1})}, 2l + 1, k_D'),$ and that the goal is to produce an estimate for $k_D',$ the starting location of the block of deletions. Suppose that the $(2l+1)$-repeating pattern identified in the above analysis starts at position $k_D''$ in $\bfx^{(1,2^{j-1})}$. Since any $(2l+1)$-repeating pattern which appears as a substring in $\bfx$ has length less than $b^4 \log n$ according to Claim~\ref{cl:reppat}, $|k_D' - k_D'' | < b^4 \log n$. Then, $\hat{k}_D = 1 + 2^{j-1} \, (k_D'' -1)$ satisfies $|\hat{k}_D - k_D| < b^5 \log n + b$. Therefore, since $\bfx \in \cC^{odd}_{\tilde{b}}(n, a_j, \bfC_j, \bfD_j)$, we can recover $\bfx$ from $\bfy$ by using the constraint $\bfx^{(i_1,t)} \in SVT_{{c_{i_1,t}},{d_{i_1,t}}}( \lfloor \frac{n-i_1}{t} \rfloor +1, 2(b^5 \log n +b)),$ along with the information about $\hat{k}_D$.
\end{IEEEproof}

The next corollary summarizes the results of this section. 

\begin{corollary}\label{cor:generalBRed} Let $\bfx, \bfu \in \cC_b(n, \bfa, \vec{\bfC}, \vec{\bfD}) \subseteq \mathbb{F}_2^n$ where $\bfx \neq \bfu$ and let $t$ be a positive integer such that $t \leq b$ for an odd positive integer $b \geq 5$. Then, $\cB_{D,t}(\bfx) \cap \cB_{D,t}(\bfu) = \emptyset$, and for any $n \geq 50b$, the code $\cC_b(n, \bfa, \vec{\bfC}, \vec{\bfD})$ introduces at most
\begin{align*}
& \lceil \log b \rceil \, \Big ( \log(bn + b^2) + \frac{b(b+1)}{2}\left( \log ( 2(b^5 \log n+b)) + 1\right) \Big ) + 1
\end{align*}
bits of redundancy.
\end{corollary}
The proof of the result proceeds along the same lines as that of Corollary~\ref{cor:oddb} and may be found in Appendix~\ref{app:generalBRed}.

\begin{remark} It is tedious, but conceptually simple, to extend the results for a single deletion and multiple adjacent transposition errors for the case of multiple deletions and multiple adjacent transpositions, even for the case of a non-binary alphabet. The key idea is to replace VT-like codes with binary codes constructed in~\cite{helberg2002multiple} and the extensions of the construction over larger fields, as presented in~\cite{le2015new}. In the former case, the VT-type constraints are replaced by what the authors refer to as number-theoretic constraints of the form
\begin{align*}
\sum_{i=1}^{n}\, v_i\,x_i=a\, \mod \, u,
\end{align*}
where the weights $v$ are defined recursively according to the formula
$$ v_j=1+\sum_{i=1}^{s}\,v_{j-i}, \;\, v_i=0, \, \forall \, i\leq 0,$$
and
$$u=1+\sum_{i=0}^{s-1}\, v_{n-i}.$$
Note that for $s \geq 2$, the codes constructed using this approach have redundancy linear in $n$.
\end{remark}

\section{Codes for Correcting an Adjacent Block Transposition and a Burst Deletion}\label{sec:blockTransDel}

Next, we describe how to construct codes capable of correcting a single block transposition along with a single block deletion. For simplicity, we limit our attention to the case where the adjacent block transposition and the block deletion are both of the same size. Furthermore, we restrict our proofs to the case of nonoverlapping bursts of deletions and transpositions, respectively. All results can be easily modified to account for this case and are omitted for clarity and compactness of exposition.

Similar to what was done in the previous section, we first outline the high level ideas behind the construction and the proof. We start with the special case when the block transposition and the block deletion are non-overlapping. Recall that for $r=n/b$ (where we tacitly assume that $b$ divides $n$), it is convenient to represent the codeword $\bfx = (x_1, \ldots, x_{n} )$ in the following manner:
\begin{align}\label{eq:cwmatrix}
\begin{bmatrix}
    x_{1}       & x_{b+1} & x_{2b+1} & \dots & x_{r(b-1) + 1} \\
    x_{2}       & x_{b+2} & x_{2b+2} & \dots & x_{r(b-1) + 2} \\
    \hdotsfor{5} \\
    x_{b}       & x_{2b} & x_{3b} & \dots & x_{n}
\end{bmatrix}.
\end{align} 
Suppose that $\bfy$ is the result of one adjacent block transposition and a block of deletions, both of length $b$. 
Note that the block deletion and adjacent block transposition have the equivalent effect of deleting one symbol from each row in the matrix representation of the codeword and swapping two adjacent symbols within each row.

One naive approach for constructing codes capable of correcting an adjacent block transposition and a block deletion is to use a $1$-TD code on each of the $b$ interleaved sequences in the matrix (\ref{eq:cwmatrix}). Since a $1$-TD code requires roughly $2 \log n$ bits of redundancy, this approach would result in a total redundancy of roughly $2b \log n$ bits. In what follows, we describe a more involved approach that requires $O( \log b \log n + b^2 \log b \log \log n)$ bits of redundancy.

The proposed construction works as follows. We first ignore the adjacent block transposition and attempt to correct the block deletion using the method of the previous section. Clearly, with this approach we may (and will) perform erroneous correction. However, the ``miscorrection'' will have a specific structure which may be exploited in the next step by using Tensor Product codes~\cite{wolf1965codes}, to be described in this section. 

To this end, we introduce the following notation. For a given word $\bfx \in \mathbb{F}^n_2$, let $\cB_{BT,b}(\bfx)$ denote the set of words that may be obtained from $\bfx$ via one adjacent block transposition of length $b$ (Recall from Section~\ref{sec:main} that we used $\cB_{(T,\ell)}(\bfx)$ to denote the set of words that may be obtained from at most $\ell$ adjacent transpositions in $\bfx$). The following simple example illustrates the newly introduced concept.

\begin{example}\label{ex:extrans} Let $\bfx = (1,0,0,0,0,0,1,1,0) \in \mathbb{F}_2^{9}$. Here, 
\begin{align*}
\cB_{BT,3}(\bfx) = \{&  (1,0,0,0,0,0,1,1,0),  (0,0,0,1,0,0,1,1,0), \\
&(1, 0,0,1,0,0,0,1,0), (1,0,0,1,1,0,0,0,0) \}.
\end{align*}
\end{example}

Recall from Section~\ref{sec:main} that with respect to the size of the relevant error balls, the order in which a single adjacent transposition 
and a single deletion occur does not matter. The next example shows that, unfortunately, this property does not carry over to the case of adjacent block transpositions and block deletions.

\begin{example}\label{ex:BTD} Let $\bfx =  (1,0,0,0,0,0,1,1,0) \in \mathbb{F}_2^{9}$. Then,  $({1,1,0},{1,0,0}) \in \cB_{BT, 3}( \cB_{D, 3}( \bfx ))$, but $({1,1,0},{1,0,0}) \not \in \cB_{D,3}( \cB_{BT,3}(\bfx))$. Similarly, let $\bfy = (1,0,$ $1,0,0,$ $0,1,1,0) \in \mathbb{F}_2^9.$ Then, 
$(1,0,0,0,0,0) \in\cB_{D,3}( \cB_{BT,3}(\bfy))$, but at the same time, $(1,0,0,0,0,0) \not \in \cB_{BT,3}( \cB_{D,3}(\bfy))$.
\end{example}
 
We would like to design codes that can correct block errors in either of the two orders, i.e., codes that can correct a block deletion followed by an adjacent block transposition \emph{and simultaneously} correct an adjacent block transposition followed by a block deletion. Hence, we need to introduce one more notion of a set, which for a word $\bfx \in \mathbb{F}_2^n$ equals
\begin{align*}
\cB_{BT \land D, b}(\bfx) = \bigcup_{t \leq b} \cB_{BT, t}( \cB_{D,t}(\bfx)) \cup \cB_{D,t}( \cB_{BT,t}(\bfx)).
\end{align*}

We have the following useful claim.

\begin{claim}\label{cl:eqburst1} For $\bfx \in \mathbb{F}_2^n$, 
$$\cB_{BT \land D,b}(\bfx) \subseteq \cB_{D, \leq b}(\cB_{(T, 2b^2)}(\bfx)).$$
\end{claim}

Using Claims~\ref{cl:trans} and~\ref{cl:eqburst1}, we can prove the following result.

\begin{corollary}\label{cor:blocktrans} Suppose that $\bfy=(y_1, \ldots, y_n) \in \cB_{(T, 2b^2)}(\bfx)$ where $\bfx \in \mathbb{F}_2^n$. Then,
$| \sum_{i=1}^n i  x_i - \sum_{i=1}^n i  y_i | \leq 2b^2.$
\end{corollary}

The claims above allows us to generalize some of the results of Section~\ref{sec:main}. To this end, recall from the previous section that for $\bfy \in \mathbb{F}_2^{n-b}$, $\bfv \in \mathbb{F}_2^b$ and $k_I \in [n-b+1]$, we used $I(\bfy, \bfv, k_I) \in \mathbb{F}_2^n$ to denote the vector obtained by inserting $\bfv$ into $\bfy$ at position $k_I$.

\begin{claim}\label{cl:vtBurstTD} Let $\bfv_1,\bfv_2 \in \mathbb{F}_2^b$. Furthermore, suppose that $\bfy \in \mathbb{F}_2^{n-b}$, $w_1=wt(\bfv_1)=wt(\bfv_2)$, and $i_1<i_2$. Let ${\bfx} \in \cB_{(T, 2b^2)}(I(\bfy, \bfv_1, i_1)) \in \mathbb{F}_2^n$, $w_2 = wt(x_{i_1+b}, \ldots, x_{i_2+b-1})$, and $\bfu \in \cB_{(T, 2b^2)}(I(\bfy, \bfv_2, i_2)) \in \mathbb{F}_2^n$. Then,
\begin{align*}
\sum_{i=1}^n i  u_i - \sum_{i=1}^n i  x_i = (i_2-i_1)  w_1 - b  w_2 + \delta + \theta,
\end{align*}
where $|\delta| < b^2$ and $|\theta|\leq 4b^2$.
\end{claim}
\begin{remark} The correction term $\theta$ arises as a consequence of allowing at most $2b^2$ adjacent transpositions to occur. The statement in Claim~\ref{cl:vtBurstTD} then follows from Corollary~\ref{cor:blocktrans}. \end{remark}

As mentioned at the beginning of the section, we first attempt to correct the block deletion. Our approach to correcting the block of deletions will be similar to that described in the previous section, where we used VT-like codes combined with coding constraints needed to approximately estimate the weight and the location of the block of deletions. Afterwards, Shifted VT codes will be used to attempt to accurately correct the deletions given the approximate starting location.

We first focus on the behavior of a single Shifted VT decoder. Recall from the previous section that $D(\bfx, b, k_D)$ is the result of deleting $b$ consecutive bits from $\bfx$ starting at position $k_D$. Furthermore, let 
$$\bfy = T(\bfx, k_T) = (x_1, \ldots, x_{k_T-1}, x_{k_T+1}, x_{k_T}, x_{k_T+2}, \ldots, x_n)$$ 
denote the word obtained by performing one adjacent transposition in $\bfx$ starting at position $k_T$. 
The next lemma characterizes the behavior of a Shifted VT decoder when it is provided with a vector that has experienced a single deletion along with a single adjacent transposition. The proof of the result, which may be found in Appendix~\ref{App:SVTProp}, follows along the same lines as that of Lemma~\ref{lem:miserror} and Corollary~\ref{cor:deltrans11}. In what follows, for a vector $\bfx \in \mathbb{F}^n_2$, we let $\rho(\bfx)$ stand for the length of the longest run of zeros or ones in $\bfx$.

\begin{lemma}\label{lem:svttdec}  Suppose that $\bfx \in SVT_{c,d}(n, P+\rho(\bfx)+2),$ where $c \in \mathbb{Z}_{P + \rho(\bfx) + 2}$, $d \in \mathbb{F}_2$, $\bfy \in D(T(\bfx, k_T),1,k_D)$, and assume that we are given a $\hat{k}_D$ such that $|\hat{k}_D-k_D| < P$.  Then, there exists a decoder $\cD_{SVT}$ for $SVT_{c,d}(n, P + \rho(\bfx) + 2)$ that can generate a vector $\bfz=I(\bfy, d_b, k_D') \in SVT_{c,d}(n, P + \rho(\bfx) + 2)$ for $d_b \in \mathbb{F}_2$ given $\bfy$ and $\hat{k}_D,$ such that $\bfz \in \cB_{(T,2)}(\bfx)$ and $|k_D'- {k}_D| < \rho(\bfx) + P$. 
\end{lemma}
Thus, similar to what we observed in the previous section, if a Shifted VT decoder is provided with a sufficiently accurate estimate of the location of the deletion, the decoder will either correct the deletion or introduce a miscorrection in the form of an additional transposition.

Our aim is to apply the result of Lemma~\ref{lem:svttdec} to each interleaved sequence, which requires all the required coding constraints - such as the VT-type constraints, balancing and runlength properties, to hold for each interleaved sequence. The individual code components are consequently integrated using tensor product codes. To allow for proper operation of the product codes, one has to ensure that the Hamming errors resulting from the deletion miscorrection stage are not ``scattered around'' but rather concentrated in terms of their locations.

We first focus on the case where the blocks have odd length, and then extend it to the general case. The ideas behind the proofs represent a combination of the approaches presented in Section~\ref{sec:main} and \ref{sec:block}. 

Let $a \in \mathbb{Z}_{bn + 5 b^2}$, and suppose that $\bfC$, $\bfD$ are defined as in (\ref{eq:oddburst}). We start by introducing the following code:

\begin{align}\label{eq:tdburst1}
\cC^{(1)}_{TD,b}(n&,a, \bfC, \bfD) = \Big \{ \bfx \in \mathbb{F}_2^n : \sum_{i=1}^n i  x_i \equiv a \bmod \Big(bn + 5 b^2 \Big),  \nonumber \\
&\bfx \in Bal(n,b), \text{ and } \forall i_2 \in [b], \forall i_1 \leq i_2,  \nonumber \\  
&\bfx^{(i_1,i_2)} \in SVT_{{c_{i_1,i_2}},{d_{i_1,i_2}}}(\lfloor \frac{n-i_1}{i_2} \rfloor +1, 2b^4 \log n+2)
\textcolor{black}{,} \nonumber  \\
& \rho(\bfx^{(i_1,i_2)}) \leq b^4 \log n \Big \}.
\end{align}

Also, let
\begin{align*}
\cC_{TD,b}^{(1)}(n,a) := \{ \bfx \in \mathbb{F}_2^n :& \sum_{i=1}^n i  x_i \equiv a \bmod \Big( bn + 5b^2 \Big), \\
&\bfx \in Bal(n,b) \}.
\end{align*}

Observe that $\cC^{(1)}_{TD,b}(n,a,\bfC, \bfD) \subseteq \cC_{TD,b}^{(1)}(n,a)$. As formally asserted in Lemma~\ref{lem:burstVT2}, the code component $\cC_{TD,b}^{(1)}(n,a)$ is used to approximately determine the \emph{location} of the odd-length burst of deletions.  The code $\cC^{(1)}_{TD,b}(n,a, \bfC, \bfD)$ is then used to attempt to correct the actual odd-length burst of consecutive deletions. As mentioned earlier, one will need another constraint (derived from tensor product codes) to correct any miscorrections introduced after this step. Note that the last code constraint in $\cC^{(1)}_{TD,b}(n,a,\bfC, \bfD)$ bounds the longest runlength of any interleaved sequence, hence ensuring that Lemma~\ref{lem:svttdec} can be invoked for each such sequence.

We formally show next that one can approximately determine the location of the block of deletions given $\cC^{(1)}_{TD,b}(n,a)$. In this context, the next lemma is an analogue of Lemma~\ref{lem:burstVT}, and its proof is given in the Appendix. 

\begin{lemma}\label{lem:burstVT2}  Let $\bfv_1,\bfv_2 \in \mathbb{F}_2^{t}$, $wt(\bfv_1) = wt(\bfv_2)$, $\bfy \in \mathbb{F}_2^{n-t}$, and suppose that $t \leq b$ is an odd number such that $b \geq 6$.  For $i_2-i_1 \geq b^4 \log n$, let ${\bfx} \in  \cB_{(T, 2b^2)}(I(\bfy, \bfv_1, i_1))$, $\bfx \in \cC^{(1)}_{TD,b}(n,a)$ and $\bfz \in \cB_{(T, 2b^2)}(I(\bfy, \bfv_2, i_2)) \in \mathbb{F}_2^n$. Then,
\begin{align*}
\sum_{i=1}^n i  z_i - \sum_{i=1}^n i  x_i \not \equiv 0 \bmod \Big(bn + 5b^2 \Big),
\end{align*}
and so $\bfz \not \in \cC^{(1)}_{TD,b}(n,a)$.
\end{lemma}

Next, we define the code $\cC^{Odd,B}_{b}(n,a, \bfC, \bfD)$, which can correct any single block deletion and adjacent block transposition when the length of the blocks is odd. In order to define the code, we need to introduce tensor product codes. 
The following definition is adapted from~\cite{gabrys2013graded}.

\begin{definition}\label{def:tpcodes} Given positive integers $t_1$ and $t_2$, a binary error vector $\bfe = (\bfe_1, \bfe_2, \ldots, \bfe_n) \in \mathbb{F}_2^{mn}$ is called an $(n,m; t_1, t_2)$ error vector if each subvector $\bfe_i,\, 1 \leq i \leq n,$ is of length $m$, and
\begin{enumerate}
\item $| \{i : {\bfe}_i \neq {\bf0} \}| \leq t_1$, and 
\item $\forall i, wt(\bfe_i) \leq t_2$.
\end{enumerate}
\end{definition}

We refer to a code $\cC \subseteq \mathbb{F}_2^{mn}$ that is capable of correcting any $(n,m; t_1, t_2)$ error vector as an \emph{$\cC(n, m ; t_1, t_2)$ code}. Suppose now that $(4 b^5 \log n + 2b)$ divides $n$. 

We define the code $\cC^{Odd,B}_{b}(n,a, \bfC, \bfD)$ according to:
\begin{align}\label{eq:oddburstBTD}
\cC^{Odd,B}_{b}(n,a,& \bfC, \bfD) = \Big \{ \bfx \in \mathbb{F}_2^n : \bfx \in \cC^{(1)}_{TD,b}(n,a, \bfC, \bfD), \nonumber\\
& \bfx \in \cC\left( \frac{n}{ 4b^5 \log n + 2b}, 4 b^5 \log n + 2b; 4, 4b \right) \Big \}.
\end{align}
The code combines two components: The first component, $\cC^{(1)}_{TD,b}(n,a, \bfC, \bfD)$, is used to perform approximate correction of a burst of consecutive deletions. Potential miscorrections introduced in the first step are corrected using the code $\cC( \frac{n}{ 4b^5 \log n + 2b}, 4 b^5 \log n + 2b; 4, 4b )$.

We have the following theorem, which relies on the result of Lemma~\ref{lem:burstVT2}. 

\begin{lemma}\label{lem:oddburstTD} Suppose that $\bfx \in \cC_{b}^{Odd,B}(n,a,\bfC, \bfD)$ and that $\bfy \in \cB_{BT \land D, b}(\bfx), \bfy \in \mathbb{F}_2^{n-t},$ where $t$ is an odd integer such that $t \leq b$ and $b \geq 6$. 
Then, there exists a decoder for $\cC_{b}^{Odd,B}(n,a,\bfC, \bfD)$ that can recover $\bfx$ from $\bfy$. \end{lemma}

\begin{IEEEproof} Since $\bfy \in \cB_{BT \land D,b}(\bfx)$, we know from Claim~\ref{cl:eqburst1} that $\bfy \in \cB_{D, \leq b}(\cB_{(T, 2b^2)}(\bfx))$. Therefore, there exists a vector $\bfv_1 \in \mathbb{F}_2^t$ and an index $k_D \in [n-t+1]$ such that $\bfx \in \cB_{(T,2b^2)}(I(\bfy, \bfv_1, k_D))$ and $\bfx \in \cC^{Odd,B}_{b}(n,a,\bfC, \bfD)$. 

Note that from the Shifted VT code constraints, we can determine $wt(\bfv_1)$ and hence produce a vector $\bfv_2$ with $wt(\bfv_1) = wt(\bfv_2)$. We then proceed to identify a vector $\bfz \in \cB_{(T,2b^2)}(I(\bfy, \bfv_2, \hat{k}_D))$ such that $\bfz \in \cC_{TD,b}^{(1)}(n,a)$ (for this purpose, one can resort to a brute force search). According to Lemma~\ref{lem:burstVT2},  for any such $\bfz$, we have $|k_D - \hat{k}_D|<b^4 \log n$.

For each $\bfy^{(i, t)},$ where $i \leq t$, we have $\bfy^{(i,t)} \in \cB_{(T,1), D}(\bfx^{(i,t)})$. Suppose that $\bfy^{(i,t)} = D(T(\bfx^{(i,t)}, k_{T,i}), 1, k_{D,i})$ and that $k_{D,i} > k_{T,i} + 1$ (The case $k_{D,i} \leq k_{T,i} + 1$ can be analyzed similarly). 
Let $\bfs^{(i,t)} = D(\bfx^{(i,t)}, k_{D,i}) \in \cB_D(\bfx^{(i,t)})$.

We use the decoder described in Lemma~\ref{lem:svttdec} to produce a vector $\bfw^{(i,t)},$ given the estimate $\hat{k}_{D,i} = \lceil \hat{k}_D/t \rceil$, and the vector $\bfy^{(i,t)}$. Suppose that $\bfx^{(i,t)} = I(\bfs^{(i, t)},d_i,k_{D,i})$. Clearly, $|\hat{k}_{D,i} -  k_{D,i} | < b^4 \log n$. According to Lemma~\ref{lem:svttdec}, $\bfw^{(i,t)}=I(\bfy^{(i,t)}, d_i, k'_{D,i}) \in \cB_{(T,2)}(\bfx^{(i,t)}),$ where $|k'_{D,i} - {k}_{D,i}| < 2b^4 \log n$. This follows since $|\hat{k}_{D,i} -  k_{D,i} | < b^4 \log n = P$, where $P$ is as described in Lemma~\ref{lem:svttdec}, and  $\rho(\bfx^{(i_1,i_2)}) \leq b^4 \log n$, which together imply that $|k'_{D,i} - {k}_{D,i}| < P + \rho(\bfx^{(i_1,i_2)}) = 2b^4 \log n$.

As a result, the miscorrections for each $\bfw^{(i,t)}$ caused by reinserting the deleted bits into the wrong locations lie close to each other; more precisely, the miscorrections are close to the position $k_D$. Hence, one may treat the miscorrections as a burst of substitutions errors that may be corrected using tensor product codes with appropriately chosen parameters.

Next, let $\bfu^{(i,t)} = T( \bfw^{(i,t)}, k_{T,i}),$ where as before,
$T(\bfx, k)$ denotes the word obtained by applying one adjacent transposition starting at position $k$ in $\bfx$. More precisely, $\bfu^{(i,t)}$ is the result of correcting the adjacent transposition that originally occurred in $\bfx^{(i,t)}$. Observe that $\bfu^{(i,t)}= I(\bfs^{(i,t)}, d_i, k'_{D,i})$, which implies that $\bfu^{(i,t)} \in \cB_{(T,1)}(\bfx^{(i,t)}),$ since $\bfw^{(i,t)} \in \cB_{(T,2)}(\bfx^{(i,t)})$ (i.e., $\bfu^{(i,t)}$ contains the miscorrections which arose from attempting to correct the adjacent transposition and deletion in $\bfx^{(i,t)}$ that lead to $\bfy^{(i,t)}$). Furthermore, since $\bfx^{(i,t)} = I(\bfs^{(i, t)},d_i,k_{D,i})$, $\bfu^{(i,t)} \in \cB_{(T,1)}(\bfx^{(i,t)})$, and $|k_{D,i} - k'_{D,i}| < 2 b^4 \log n$, there exists a $k'_{T,i}$ such that $|k'_{T,i} - k_{D,i}| < 2 b^4 \log n + 1$ and $\bfx^{(i,t)} = T(\bfu^{(i,t)}, k'_{T,i})$, so that we can correct the error in $\bfu^{(i,t)}$ by transposing two symbols in $\bfu^{(i,t)}$ that are within distance $2b^4 \log n + 1$ from the position $k_{D,i}$.

Thus, we have at most two pairs of mismatched symbols between $\bfx^{(i,t)}, \bfw^{(i,t)}$, which appear at positions $i_1, i_2$ and $j_1, j_2$ in $\bfx$ and $\bfw$, respectively. Suppose, without loss of generality, that the pair of errors in $\bfw^{(i,t)}$ at positions $i_1, i_2$ in $\bfw$ are due to the adjacent block transposition and that the pair of errors in $\bfw^{(i,t)}$ at positions $j_1, j_2$ in $\bfw$ are due to the miscorrections associated with the Shifted VT decoders. Then, $|k_D - j_1| < 2b^5 \log n + b$, $|k_D - j_2| < 2b^5 \log n + b$. Using the same arguments for $\bfx^{(i,t)}, \bfw^{(i,t)}$ where $1 \leq i \leq b$, we conclude that $\bfw$ and $\bfx$ differ by at most a $(\frac{n}{4 b^5 \log n+2b}, 4 b^5 \log n + 2b; 4, 4b)-$type error. Since $\bfx$ belongs to a $\cC(\frac{n}{4 b^5 \log n+2b},4 b^5 \log n + 2b; 4, 4b)-$error correcting code, the claimed result follows.
\end{IEEEproof}

\begin{remark} Note that when $(4b^5 \log n + 2b) \not| \; n$, one can use the same approach as the one described in the previous lemma with a tensor product code of length $(4b^5 \log n + 2b) \cdot \lceil \frac{n}{4b^5 \log n + 2b} \rceil$. Here, where we assume that the last $(4b^5 \log n + 2b) \cdot \lceil \frac{n}{4b^5 \log n + 2b} \rceil -n$ positions of the tensor product code of length $(4b^5 \log n + 2b) \cdot \lceil \frac{n}{4b^5 \log n + 2b} \rceil$ are set to zero. More precisely, we replace the condition $\bfx \in \cC\left( \frac{n}{4b^5 \log n + 2b}, 4b^5 \log n + 2b; 4, 4b \right)$ in (\ref{eq:oddburstBTD}) with $\tilde{\bfx} \in \cC\left( \lceil \frac{n}{4b^5 \log n + 2b} \rceil, 4b^5 \log n + 2b; 4, 4b \right)$, where $\bfx$ and $\tilde{\bfx}$ agree in the first $n$ positions and where $\tilde{\bfx}$ is set to zero in the remaining positions.
\end{remark}

We illustrate the encoding/decoding procedures with the following example.

\begin{example} Suppose that 
$$\bfx = (1,1,1,0,1,0,1,0,1,1,0,1,0,0,0,1,0,1,1,1,1) \in \mathbb{F}_2^{21}$$
was transmitted and that $\bfy = (0,1,0,1,1,1,1,0,1,1,$ $0,1,0,0,0,1,0,1) \in \cB_{D,3}(\cB_{BT,3}(\bfx))$ was received instead. It is straightforward to check that $\sum_{i=1}^{21} i \, x_i \equiv 35 \bmod 108$.

As the first step of decoding, we find a vector $\bfv_2 \in \mathbb{F}_2^3$ and another vector $\bfz = I(\bfy, \bfv_2, \hat{k})$ such that $wt(\bfv_2) = 3$ and $\sum_{i=1}^n i \, z_i \equiv 35 \bmod 108$. There exists only one vector $\bfz = (0,1,0,1,1,1,1,0,1,1,0,1,0,0,1,1,1,0,1,0,1)$ for which $\bfv_2 = (1,1,1)$ and $\hat{k}=15$.

Notice that $\bfx^{(1,3)} = (1,0,1,1,0,1,1)$ and that $\sum_{i=1}^7 i \, x_i^{(1,3)} \equiv 5 \bmod 8$. 
Thus, given that we know the value of the deleted bit, the parameter $\hat{k}_1=5$ (which is an estimate derived from $\hat{k},$ where $\hat{k}_1 = \lceil \frac{\hat{k}}{3} \rceil$), and $\bfy^{(1,3)} = (0,1,1,1,0,1)$, we may use the decoder described in Lemma~\ref{lem:svttdec} to generate $\bfw^{(1,3)} = (0,1,1,1,1,0,1)$. Similarly, as $\bfw^{(2,3)} = (1,1,0,0,0,0,1)$ and $\bfw^{(3,3)} = (0,1,1,1,1,0,1),$ one has $\bfw = (0,1,0,1,1,1,1,0,1,1,0,1,1,0,1,0,0,0,1,1,1)$ (We have highlighted the positions where $\bfw$ and $\bfx$ differ). We can correct the remaining errors using a $(7,3; 4,2)-$type code from Definition~\ref{def:tpcodes}.
\end{example}

The following corollary follows from Lemma~\ref{lem:oddburstTD} and the well-known Gilbert-Varshamov bound. The proof is given in Appendix~\ref{app:poddb}.

\begin{corollary}\label{cor:oddTD} Let $\cC^{Odd,B}_{b}(n, \bfa, {\bfC}, {\bfD})$ be as defined in (\ref{eq:oddburstBTD}). Let $t$ be a an odd positive integer such that $t \leq b$, with $b \geq 6$. Then, the code is a single transposition and block deletion correcting code. Furthermore, for any $n \geq 10$,
\begin{align*}
&\log | \cC^{Odd,B}_{b}(n, \bfa, {\bfC}, {\bfD}) | \geq \\
& n- [\log(bn + 5b^2) + \frac{b(b+1)}{2}\left( \log ( 2 b^4 \log n + 2) + 1\right)  \\
&+ 8 \log n + 64b \log (4 b^5 \log (n) + 2b) + 2].
\end{align*}
\end{corollary}

An immediate consequence of the above corollary is that for $b = O(1)$, one has
$$ \log | \cC^{Odd,B}_{b}(n, \bfa, {\bfC}, {\bfD}) | \geq n- [9 \log n + O(\log \log n)]. $$

We are now ready to state the general code construction using the same approach as that described in the previous section. 
In particular, the next result proves the existence of a code capable of correcting any block of deletions and an adjacent block transposition; the redundancy of the construction is approximately $\log b \, \left( \log n + O(\log \log n) \right) + 8 \log n$ bits.

Define the codebook $\cC_{TD,b}(n,\bfa, \vec{\bfC},\vec{\bfD})$ according 
\begin{align*}
\cC_{TD,b}(n,&\bfa, \vec{\bfC},\vec{\bfD}) = \Big \{ \bfx \in \mathbb{F}_2^n : \\
& \bfx \in \cC\left( \frac{n}{4 b^5 \log n + b}, 4 b^5 \log n + b; 4, 4b \right),  \\
&\forall j \in [\lceil \log b \rceil ], \bfx^{(1,2^{j-1})} \in \cC^{(1)}_{TD,\tilde{b}}(\lceil \frac{n}{2^{j-1}} \rceil , a_j, \bfC_j, \bfD_j) \\
&\text{with } \tilde{b} = \max \{ \lceil \frac{b}{2^{j-1}} \rceil, 6 \}  \Big \},
\end{align*}
with parameters $\bfa = (a_1, \ldots, a_{\lceil \log b \rceil})$, $\vec{\bfC} = (\bfC_1, \ldots, \bfC_{\lceil \log b \rceil})$, $\vec{\bfD} =( \bfD_1, \ldots, \bfD_{\lceil \log b \rceil})$. The codes $\cC^{(1)}_{TD,\tilde{b}}(\lceil \frac{n}{2^{j-1}} \rceil, a_j, \bfC_j, \bfD_j)$ are as defined in (\ref{eq:tdburst1}). 

The following theorem is a consequence of Lemma~\ref{lem:oddburstTD} and may be proved similarly as Theorem~\ref{th:burstcode}.

\begin{theorem}\label{th:burstcode2} Suppose that $\bfx \in \cC_{TD,b}(n,\bfa, \vec{\bfC}, \vec{\bfD})$ is transmitted and that $\bfy \in \cB_{BT \land D,b}(\bfx)$ is received. Then, there exists a decoder for $\cC_{TD,b}(n,\bfa, \vec{\bfC}, \vec{\bfD})$ capable of uniquely determining $\bfx$ from $\bfy$.
\end{theorem}

\begin{corollary} Let $\bfx, \bfu \in \cC_{TD,b}(n, \bfa, \vec{\bfC}, \vec{\bfD})$ be defined as above. Then, for any $n \geq 50b$, $b \geq 6$, the code is a block transposition and deletion correcting code satisfying
\begin{align*}
 &\log | \cC_{TD,b}(n, \bfa, \vec{\bfC}, \vec{\bfD}) | \geq \\
& n- [\lceil \log b \rceil \, ( \log(bn + 5b^2) + \frac{b(b+1)}{2}\left( \log ( 2 b^4 \log n + 2) + 1\right) + 2 )  \\
&+ 8 \log n + 64b \log (4 b^5 \log (n) + 2b)].
\end{align*}
\end{corollary}

\section{Conclusion}

We introduced a new family of error-correction codes termed \emph{codes in the Damerau distance}. Codes in the Damerau distance are capable of correcting deletions as well as a limited number of adjacent symbol transposition errors. Given that adjacent transpositions may be viewed as correlated pairs of deletions and insertions, Damerau codes also represent a family of codes capable of correcting both random and correlated deletion/insertion patterns. 
We proposed generalizations of VT codes that are capable of correcting one deletion or one transposition, or one deletion and one transposition error. We then proceeded to address the more challenging problem of designing codes that may correct one deletion and multiple adjacent transposition errors. Using state-of-the-art burst-deletion correcting codes described and analyzed in the paper, we also extended the aforementioned results to the case of block deletions and transpositions. 

The Summary Table highlights the main results of our work, which include the cardinalities of several new families of codes. Each row of the table represents a code introduced in the paper; the entries in the first column are descriptions of the code in terms of the types of errors that can be corrected; the entries in the second column indicate where the codes may be found in the paper, while 
the entries in the third column list the number of redundant bits required in the construction.

\begin{table*}\label{summary}
\begin{center}
\begin{tabular}{ c| c | c }
 Code Description  & Position in Manuscript & Redundancy (in bits) \\ 
\hline
A deletion or an adjacent transposition.  & Section~\ref{sec:basics} & $\log(6n - 3)$  \\
A deletion and an adjacent transposition. & Section~\ref{sec:main} & $2 \log n + \mathcal{O}(1)$ \\
A deletion and $t$ adjacent transpositions. & Section~\ref{sec:main} & $2t \log\,n + \log (n+2t+1)$ \\
A burst of at most $b$ consecutive deletions (a block deletion). & Section~\ref{sec:block} & $\log b \, \log n + \mathcal{O}(b^2 \, \log\, b \, \log \log n)$ \\ 
An adjacent block transposition and a block deletion \textcolor{black}{of constant length} & Section~\ref{sec:blockTransDel} & $9 \log n + \cO(\log \log n)$
\end{tabular} 
\end{center}
\caption {Summary of Results: Main Constructions and Redundancy.}
\end{table*}

Open problems regarding codes in the Damerau distance include describing efficient constructions that may correct multiple adjacent transposition and deletion errors, as well as multiple block deletion and adjacent transposition errors.

\section*{Author biographies}

\textbf{Ryan Gabrys} received his Ph.D. degree in electrical engineering from the University of California Los Angeles. Since 2014, he has been a postdoctoral researcher at the University of Illinois Urbana Champaign. Currently, he works at SPAWAR Systems Center San Diego. His research interests include coding theory with applications to storage and synchronization.\\

\textbf{Olgica Milenkovic} is a professor of Electrical and Computer Engineering at the University of Illinois, Urbana-Champaign (UIUC), and Research Professor at the Coordinated Science Laboratory. She obtained her Masters Degree in Mathematics in 2001 and PhD in Electrical Engineering in 2002, both from the University of Michigan, Ann Arbor. Prof. Milenkovic heads a group focused on addressing unique interdisciplinary research challenges spanning the areas of algorithm design and computing, bioinformatics, coding theory, machine learning and signal processing. Her scholarly contributions have been recognized by multiple awards, including the NSF Faculty Early Career Development (CAREER) Award, the DARPA Young Faculty Award, the Dean's Excellence in Research Award, and several best paper awards. In 2013, she was elected a UIUC Center for Advanced Study Associate and Willett Scholar. In 2015, she became Distinguished Lecturer of the Information Theory Society. From 2007 until now, she has served as Associate Editor of the IEEE Transactions of Communications, the IEEE Transactions on Signal Processing, the IEEE Transactions on Information Theory and the IEEE Transactions on Molecular, Biological and Multi-Scale Communications. In 2009, she was the Guest Editor in Chief of a special issue of the IEEE Transactions on Information Theory on Molecular Biology and Neuroscience.\\

\textbf{Eitan Yaakobi} (S'07--M'12--SM'17) is an Assistant Professor at the Computer Science Department at the Technion Israel Institute of Technology. He received the B.A. degrees in computer science and mathematics, and the M.Sc. degree in computer science from the Technion Israel Institute of Technology, Haifa, Israel, in 2005 and 2007, respectively, and the Ph.D. degree in electrical engineering from the University of California, San Diego, in 2011. Between 2011-2013, he was a postdoctoral researcher in the department of Electrical Engineering at the California Institute of Technology. His research interests include information and coding theory with applications to non-volatile memories, associative memories, data storage and retrieval, and voting theory. He received the Marconi Society Young Scholar in 2009 and the Intel Ph.D. Fellowship in 2010-2011.

\bibliographystyle{IEEEtranS}
\bibliography{damerau-ref}

\begin{appendices}

\section{Proof of Claim~\ref{cl:CSB}}\label{App:clCSB}

We first evaluate the probability that the first $M=b^4 \log n$ bits of $\bfx$ have less than $M/2 - (M/3b)$ or more than $M/2 + (M/3b)$ ones. Let $\bfx$ be a uniformly at random selected element from $\mathbb{F}_2^n$. For any $i \in [n]$, let $X_i$ be the indicator random variable that takes the value one when $x_i=0$ and zero otherwise. Then, $(X_1, \ldots, X_M)$ is an i.i.d random vector over $\{0,1\}$. Invoking Hoeffding's inequality we obtain 

$$ P\left( \sum_{i=1}^M x_i \geq \frac{M}{2} + \frac{M}{3b}\right) = P\left( \sum_{i=1}^M x_i \geq \frac{M}{2} - \frac{M}{3b}\right)\leq e^{-\frac{2M}{9b^2}}.$$
Let 
$$f(M,b) = e^{-\frac{2M}{9b^2}}.$$ 
Note that $f(M,b)$ is decreasing in $M$, since 
$$\frac{\partial f(M,b)}{\partial M} = - \frac{2 e^{-\frac{2M}{9b^2}}}{9 b^2}.$$ 
Applying the union bound leads to
\begin{align*}
P(\bfx \not \in Bal(n,b) ) \leq 2 n^2 \, f(b^4 \log n, b).
\end{align*}
Thus, $|Bal(n,b)| \geq 2^n \, \left( 1 - 2n^2 \, f( b^4 \log n, b) \right)$ and so
\begin{align*}
\log | Bal(n,b) | \geq  n + \log  \left( 1-2 \, n^{2-\frac{2}{9}b^2\log e} \right).
\end{align*}

\section{Proof of deletion capability of the Shifted VT Codes}\label{App:SVTProp}

Here, we prove that the Shifted VT codes are able to determine the location of a deletion given a sufficiently accurate estimate of the location of the deletion. Recall from the previous exposition that a Shifted VT code, denoted $SVT_{c,d}(n, P)$, is defined as:
\begin{align*}
SVT_{c,d}(n,M) = \{ \bfx \in \mathbb{F}_2^n :& \sum_{i=1}^n i \, x_i \equiv c \bmod {M}, \\
&\sum_{i=1}^n x_i \equiv d \bmod 2 \}.
\end{align*}

The next two claims are straightforward to prove.

\begin{claim} Let $\bfy \in \mathbb{F}_2^{n-1}$, $d_b \in \mathbb{F}_2$, and suppose that $\bfx = I(\bfy, d_b, i_1)$ and $\bfu = I(\bfy, d_b, i_2),$ where $i_2 > i_1$. Let $w=wt(y_{i_1}, \ldots, y_{i_2-1})$. Then, for any $k \in [n]$,
$$ \sum_{i=1}^n (k+i) u_i - \sum_{i=1}^n (k+i) x_i =  (i_2-i_1) d_b - w. $$
\end{claim}

\begin{claim}\label{cl:twistVT} Let $\bfy \in \mathbb{F}_2^{n-1}$, $d_b \in \mathbb{F}_2$, and suppose that $\bfx = I(\bfy, d_b, i_1)$ and $\bfu = I(\bfy, d_b, i_2),$ where $i_2 > i_1$ so that $|i_2 - i_1| < P$. 
Then, for any $k \in [n]$ and $M \geq P$, it holds that
$$ \sum_{i=1}^n (k+i) \, x_i \not \equiv \sum_{i=1}^n (k+i) \, u_i \bmod M, $$
unless $\bfx = \bfu$.
\end{claim}

As a consequence of the previous claim, we may prove the following lemma, which describes the deletion-correcting capabilities of Shifted VT codes.
\ \\

\noindent \textbf{Lemma 16.} Suppose that $\bfy \in D(\bfx, 1, k_D)$, where $\bfx \in SVT_{c,d}(n,M)$ and where $M \geq 2P-1$, $d_b \in \mathbb{F}_2$. Given $\hat{k}_D$ is such that $|k_D - \hat{k}_D| < P$, there exists at most one possible value for $k_D'$ and one possible value for $\bfd_b$ that jointly satisfy $I(\bfy, d_b, k_D') \in SVT_{c,d}(n,M)$. In this setting, we have $I(\bfy, d_b, k_d') = \bfx$. \\ 
\begin{IEEEproof} First, notice that we can determine the value of the bit deleted from $\bfx$ from the constraint $\sum_{i=1}^n x_i \equiv d \bmod 2,$ since $\bfx \in SVT_{c,d}(n,M)$. 

Let $d_b \in \mathbb{F}_2$ be the value of the deleted bit. Let $\bfy_1 = (y_1, \ldots, y_{\hat{k}_D - P})$ and $\bfy_2 = (y_{\hat{k}_D + P-1}, \ldots, y_{n-1})$. We have $(x_1, \ldots, x_{\hat{k}_D -P}) = (y_1, \ldots, y_{\hat{k}_D-P})$ and $ (x_{\hat{k}_D + P}, \ldots, x_{n}) =  (y_{\hat{k}_D + P-1}, \ldots, y_{n-1}),$ since $|k_D - \hat{k}_D|<P$. Let 
$$ c' \equiv \sum_{i=1}^{\hat{k}_D-P} i \, y_i + \sum_{i=\hat{k}_D + P-1}^{n-1} (i+1) \, y_i \bmod M.$$
Then
$$ \sum_{i=\hat{k}_D - P + 1}^{\hat{k}_D+P-1} i \, x_i \equiv c - c' \bmod M, $$
where $c$ is, as we recall, one of the parameters of the Shifted VT code.
Let $\bfu = (x_{\hat{k}_D-P+1}, \ldots, x_{\hat{k}_D+P-1})$ and observe that $\hat{\bfy} =  (y_{\hat{k}_D-P+1}, \ldots, y_{\hat{k}_D+P-2}) \in \cB_D(\bfu)$. Clearly, if $\bfu$ is known then $\bfx = (\bfy_1, \bfu, \bfy_2)$. After a change of variables, we obtain 
$$ \sum_{j=1}^{2P-1} (\hat{k}_D - P + j) \, u_j \equiv c - c' \bmod M.$$
According to Claim~\ref{cl:twistVT}, we can now recover $\bfu$ given the previous equation and $\hat{\bfy}$. This proves the lemma.
\end{IEEEproof}

In the following derivations, we once more make use of the vector
$$\bfy = T(\bfx, k_T) = (x_1,\ldots, x_{k_T-1},x_{k_T+1}, x_{k_T},x_{k_T+2}, \ldots, x_n).$$

\noindent \textbf{Lemma 22.} Suppose that $\bfx \in SVT_{c,d}(n, P+\rho(\bfx)+2),$ where $c \in \mathbb{Z}_{P + \rho(\bfx) + 2}$, $d \in \mathbb{F}_2$, $\bfy \in D(T(\bfx, k_T),1,k_D)$, and assume that we are given a $\hat{k}_D$ such that $|\hat{k}_D-k_D| < P$.  Then, there exists a decoder $\cD_{SVT}$ for $SVT_{c,d}(n, P + \rho(\bfx) + 2)$ that can generate a vector $\bfz=I(\bfy, d_b, k_D') \in SVT_{c,d}(n, P + \rho(\bfx) + 2)$ for $d_b \in \mathbb{F}_2$ given $\bfy$ and $\hat{k}_D,$ such that $\bfz \in \cB_{(T,2)}(\bfx)$ and $|k_D'- {k}_D| < \rho(\bfx) + P$.
\begin{IEEEproof} Suppose that $k_T +1 < k_D$ (The case $k_T + 1 > k_D$ may be proved by applying the same argument to the reverses of the sequences). 

Similarly as in the proof of Lemma~\ref{lem:SVTBasics}, let $\bfy_1 = (y_1, \ldots, y_{\hat{k}_D - P})$, $\bfy_2 = (y_{\hat{k}_D + P-1}, \ldots, y_{n-1})$, and $\bfu = (x_{\hat{k}_D-P+1}, \ldots, x_{\hat{k}_D+P-1})$. 

Also, let $\hat{\bfy}=(y_{\hat{k}_D-P+1}, \ldots, y_{\hat{k}_D+P-2})$.

First, we consider the case when $k_T \in \{ \hat{k}_D-P+1, \ldots, \hat{k}_D+P-1 \}$. Then, we have $\hat{\bfy} \in \cB_{(T,1),D}(\bfu)$. Letting $c'$ be defined as in the proof of Lemma~\ref{lem:SVTBasics}, we can show that 
$$ \sum_{j=1}^{2P-1} (\hat{k}_D - P + j) \, u_j \equiv c - c' \bmod P + \rho(\bfx) + 2,$$ 
and can hence recover the value of the deleted bit from $\sum_{j=1}^{2P-1} u_j \bmod 2$. The claimed result now follows from Corollary~\ref{cor:deltrans11}.

Next, suppose that $k_T < \hat{k}_D-P+1$. We assume that $k_D$ is not in the first or last run of the vector $\bfy$ (The case when $k_D$ is in the first or last run can be proved similarly, but is slightly more technical). Let $k_U$ be the largest index such that both $y_{k_U}=y_{k_D}$ and $y_{k_U},y_{k_D}$ belong to the same run. Similarly, let $k_L$ be the smallest index such that both $y_{k_L}=y_{k_D}$ and $y_{k_L},y_{k_D}$ belong to the same run.

Suppose that $d_b \in \mathbb{F}_2$ is the bit deleted from $\bfx$. If $x_{k_{T}}=x_{k_D}$, set $\bfz = T(I(\bfy, d_b, k_D), k_L-1)$, and notice that $\bfz = T(I(\bfy, d_b, k_D), k_L-1) = I(\bfy, d_b,k_L-1) \in SVT_{c,d}(n, P + \rho(\bfx) + 2)$ and $\bfz = T(T(\bfx, k_L-1), k_T)$. Note that $k_D - (k_L - 1) \leq \rho(\bfx)$. Since $|\hat{k}_d - k_d| < P$, it follows that $|\hat{k}_d - (k_L - 1)|< \rho(\bfx) + P$. Clearly, $\bfz \in \cB_{(T,2)}(\bfx),$ and from Lemma~\ref{lem:SVTBasics}, $\bfz=I(\bfy, d_b,k_L-1)$ is unique. The case $x_{k_{T}} \neq x_{k_D}$ may be handled similarly.
\end{IEEEproof}

\section{Proof of Corollary~\ref{cor:oddb}}\label{app:pfcoroddb}

The claim that for $\bfx, \bfu \in \cC_b^{Odd}(n,a, \bfC, \bfD) \subseteq \mathbb{F}_2^n$, one has $\cB_{D,t}(\bfx) \cap \cB_{D,t}(\bfu) = \emptyset$ follows immediately from Theorem~\ref{th:oddburst}.

Regarding the claim about the code redundancy, we consider the set of ``balanced'' words $Bal(n,b)$ as defined in (\ref{eq:codingc}) and apply an averaging argument which involves the parameters $a, c_{i_1, i_2}, d_{i_1, i_2}$. Then,
\begin{align*}
|\cC_b^{Odd}(n,a, \bfC, \bfD)| \geq \frac{|Bal(n,b)|}{(bn + b^2) \, \prod_{i_2=1}^b \prod_{i_1=1}^{i_2} 2\, (2 (b^5 \log n + b) )}.
\end{align*}
Taking the logarithms of both sides provides the claimed result.

\section{Proof of Corollary~\ref{cor:generalBRed}}\label{app:generalBRed}

The claim that for $\bfx, \bfu \in \cC_b(n, \bfa, \vec{\bfC}, \vec{\bfD}) \subseteq \mathbb{F}_2^n$, $\cB_{D,\leq b}(\bfx) \cap \cB_{D, \leq b}(\bfu) = \emptyset$ follows from Theorem~\ref{th:burstcode}. From the constraints in (\ref{eq:generalBurst}), if $\bfx \in \cC_b(n,\bfa, \vec{\bfC},\vec{\bfD})$, then for $j \in [ \lceil \log b \rceil ]$, we have 
$$\bfx^{(1,2^{j-1})} \in \cC^{Odd}_{\tilde{b}}(\lceil \frac{n}{2^{j-1}} \rceil , a_j, \bfC_j, \bfD_j),$$
where $\tilde{b} = \max \{ \lceil b/2^{j-1} \rceil, 5 \}$. Thus, $\bfx^{(1,2^{j-1})} \in Bal(\lceil n/2^{j-1} \rceil, \tilde{b})$. Clearly, from (\ref{eq:codingc}), we have $|Bal(\lceil n/2^{j-1} \rceil, \tilde{b})| \geq |Bal(\lceil n/2^{j-1} \rceil, b)|$. Invoking the proof of Claim~\ref{cl:CSB} with $b \geq 5$, and applying the union bound, we arrive at the bound
\begin{align*}
P \big (& \exists j \in [ \lceil \log b \rceil ], x^{(1, 2^{j-1})} \not \in Bal(\lceil \frac{n}{2^{j-1}} \rceil, \tilde{b}) \big)\\
&\leq \lceil \log b \rceil \, 2 \, (\frac{n}{b})^{2-\frac{2b^2}{9 \log_e(2)}}\\
&\leq \frac{1}{2}
\end{align*}
which holds whenever $n \geq 50\,b$. 
Thus, using similar arguments as those invoked in the proof of Corollary~\ref{cor:oddb}, we have
\begin{align*}
&|\cC_b(n, \bfa, \vec{\bfC}, \vec{\bfD})| \geq \\
&\frac{2^{n-1}}{\left( (bn + b^2) \, \prod_{i_2=1}^b \prod_{i_1=1}^{i_2} 2\, (2(b^5 \log n +b)) \right)^{\lceil \log b \rceil}}.
\end{align*}

\section{Proof of Lemma~\ref{lem:burstVT2}}
We repeat the same steps of the proof used to establish Lemma~\ref{lem:burstVT}. 

Let $w_1=wt(\bfv_1) = wt(\bfv_2)$ and $w_2 = wt(x_{i_1+t}, \ldots, x_{i_2+t-1})$.  Now according to Claim~\ref{cl:vtBurstTD}, 
$$\sum_{i=1}^n i \, z_i - \sum_{i=1}^n i \, x_i = (i_2-i_1) \, w_1 - t \, w_2 + C + D$$ 
and so in what follows we focus on showing that
\begin{align}\label{eq:lem22mainid}
B \, w_1 + C + D \not \equiv t \, w_2 \bmod bn + 5b^2
\end{align}
for $B = i_2 - i_1 \geq b^4 \log n$.

Since $\bfx \in \cC_{TD,b}^{(1)}(n,a)$, we have $\bfx \in Bal(n,b)$ and so 
$$\frac{B}{2} - \frac{B}{3b}< \sum_{i=i_1+t}^{i_2+t-1} x_i < \frac{B}{2} + \frac{B}{3b}$$ 
follows from (\ref{eq:codingc}). Thus, since $t\leq b$
\begin{align*}
\frac{Bt}{2} - \frac{B}{3} < t \, w_2 < \frac{Bt}{2} + \frac{B}{3}.
\end{align*}
Notice that since $t$ is odd, and since $w_1 = (t+k)/2$, $-b \leq k \leq b$, $k$ has to be odd. Thus, we have
\begin{align*}
B w_1 + C + D = \frac{Bb}{2} + k \, \frac{B}{2} + C + D,
\end{align*}
where $k \neq 0$. We will prove the result for the case when $k$ is positive (The case when $k$ is negative may be proved using the same argument). For $k \geq 1$, we have 
$$B w_1 + C + D \geq \frac{Bt}{2} + \frac{B}{2} - b^2 - 4b^2.$$ 
Since $B \geq b^4 \log n$ and $b \geq 6$, one has 
$$ B w_1 + C + D \geq \frac{Bt}{2} + \frac{B}{2} - 5 b^2 > \frac{Bt}{2} + \frac{B}{3} > t \, w_2, $$
so that (\ref{eq:lem22mainid}) holds.

\section{Proof of Corollary~\ref{cor:oddTD}}\label{app:poddb}

From~(\ref{eq:oddburstBTD}), we may write
\begin{align*}\
\cC^{Odd,B}_{b}(n,a,& \bfC, \bfD) = \Big \{ \bfx \in \mathbb{F}_2^n : \bfx \in \cC^{(1)}_{TD,b}(n,a, \bfC, \bfD), \nonumber\\
& \bfx \in \cC\left( {\frac{n}{4 b^5 \log n + 2b}}, {4 b^5 \log n + 2b}; 4, 4b \right)  \Big \}.
\end{align*}

Recall that $\cC^{(1)}_{TD,b}(n,a,\bfC, \bfD)$ is such that
\begin{align*}
\cC^{(1)}_{TD,b}(n&,a, \bfC, \bfD) = \Big \{ \bfx \in \mathbb{F}_2^n : \sum_{i=1}^n i  x_i \equiv a \bmod \Big(bn + 5 b^2 \Big),  \nonumber \\
&\bfx \in Bal(n,b), \text{ and } \forall i_2 \in [b], \forall i_1 \leq i_2,  \nonumber \\  
&\bfx^{(i_1,i_2)} \in SVT_{{c_{i_1,i_2}},{d_{i_1,i_2}}}(\lfloor \frac{n-i_1}{i_2} \rfloor {+1}, {2b^4 \log n+2}), \nonumber  \\
& {\rho(\bfx^{(i_1,i_2)}) \leq b^4 \log n} \Big \}.
\end{align*}

Let $\cY$ denote the following set
\begin{align*}
\cY = \Big \{& \bfx \in \{0,1\}^n : \bfx \in Bal(n,B), \\
&\forall i_2 \in [b], \forall i_1\leq i_2, \rho(\bfx^{(i_1, i_2)}) \leq b^4 \log n \Big \}.
\end{align*}

Using the union bound along with Claim~\ref{cl:CSB}, we have
\begin{align*}
|\cY| \geq 2^n  - b^2 \cdot n \cdot 2^{n-b^4 \log n} - 2^{n-1} \geq 2^{n-2}
\end{align*}
for $n \geq 10$ and $b \geq 6$. Repeating the same arguments as invoked in Corollary~\ref{cor:oddb}, we have 
\begin{align*}
|\cC^{(1)}_{TD,b}(n,a, \bfC, \bfD)| \geq \frac{|\cY|}{(bn + 5b^2) \, \prod_{i_2=1}^b \prod_{i_1=1}^{i_2} 2\, (2 b^4 \log n+2 )},
\end{align*}
and so 
$$n - \log |\cC^{(1)}_{TD,b}(n,a, \bfC, \bfD)| \leq \log (bn+5b^2) + \frac{b(b+1)}{2}$$ 
$$\left( \log(2b^4 \log n + 2) + 1 \right) + 2.$$

The parity check matrix of a $\cC( \frac{n}{4 b^5 \log n + 2b}, 4 b^5 \log n + 2b; 4, 4b )-$type code can be formed as follows~\cite{gabrys2013graded}. Let $H_2$ be a parity-check matrix of a binary code $\cC_2$ with Hamming distance $8b+1$ and of length $4\,b^5 \log n + 2b$. 
Also, let $H_{q}$ be a parity-check matrix of a non-binary code $\cC_q$ over $\mathbb{F}_q$ that has minimum Hamming distance $9$ and length $n/(4 b^5 \log n + 2b)$. Then a parity-check matrix for a $\cC(n/(4 b^5 \log n + 2b), 4 b^5 \log n + 2b; 4, 4b)-$type code can be formed by taking the tensor product $H_q \otimes H_2$.

Applying the Gilbert-Varshamov bound, we obtain $n-\log |\cC_2| \leq 8b \log \left( 4 b^5 \log n + 2b \right)$ and so 
$$n-\log |\cC_q| \leq 64 b \log \left( 4 b^5 \log n + 2b \right) + 8 \log n.$$  Using the same averaging arguments as before establishes the claim in the corollary.

\end{appendices}
\end{document}